\documentclass[aps,showkeys,singlespace,showpacs,preprintnumbers,amsmath,amssymb,prb]{revtex4}

\usepackage{graphicx} 
\usepackage{dcolumn} 
\usepackage{bm} 
\usepackage{amsmath}

\usepackage{color}
\usepackage{subfigure}

\newcommand{\efunc}[1]{\ensuremath{\mathrm{e}^{#1}}}
\newcommand{\avg}[2]{\ensuremath{{\left\langle{#2}\right\rangle}_{#1}}}
\newcommand{\muex}[1]{\ensuremath{\mu^{\mathrm{ex}}_{\mathrm{#1}}}}

\newcommand{\pd}[2]{\ensuremath{\frac{\partial {#1}}{\partial {#2}}}}

\begin{document}

\title{Quasi-Chemical and Structural Analysis of Polarizable \\ Anion Hydration}

\author{David M. Rogers\footnote{Present address: Sandia National Laboratories, Albuquerque, NM 87185} and Thomas L. Beck}
\email{thomas.beck@uc.edu} 
\affiliation{Department of Chemistry \\ University of Cincinnati \\ Cincinnati, OH 45221-0172}


\date{\today}

\begin{abstract}
Quasi-chemical theory is utilized to analyze the relative roles of solute polarization and size in determining the structure and thermodynamics of bulk anion hydration for the Hofmeister series Cl$^-$, 
Br$^-$, and I$^-$.  Excellent agreement with experiment is obtained for whole salt hydration free energies using the polarizable AMOEBA force field.  The total hydration free energies display a stronger dependence on ion size than on polarizability. The quasi-chemical approach exactly partitions the solvation free energy into inner-shell, outer-shell packing, and outer-shell long-ranged contributions by means of a hard-sphere condition. The inner-shell contribution becomes slightly more favorable with increasing ion polarizability, indicating electrostriction of the nearby waters.  Small conditioning radii, even well inside the first maximum of the ion-water(oxygen) radial distribution function, result in Gaussian behavior for the long-ranged contribution that dominates the ion hydration free energy.  This in turn allows for a mean-field treatment of the long-ranged contribution, leading to a natural division into first-order electrostatic, induction, and van der Waals terms.  The induction piece exhibits the strongest ion polarizability dependence, while the larger-magnitude first-order electrostatic piece yields an opposing but weaker polarizability dependence.  The van der Waals piece is small and positive, and it displays a small ion specificity. The sum of the inner-shell, packing, and long-ranged van der Waals contributions exhibits little variation along the anion series for the chosen conditioning radii, targeting electrostatic effects (influenced by ion size) as the largest determinant of specificity.  In addition, a structural analysis is performed to examine the solvation anisotropy around the anions.  As opposed to the hydration free energies, the solvation anisotropy depends more on ion polarizability than on ion size: increased polarizability leads to increased anisotropy.  The water dipole moments near the ion are similar in magnitude to bulk water, while the ion dipole moments are found to be significantly larger than those observed in quantum mechanical studies.  Possible impacts of the observed over-polarization of the ions on simulated anion surface segregation are discussed.  
\end{abstract}

\maketitle

\section{ Introduction}
\label{sec:intro}


Specific ion or Hofmeister effects are exhibited in a wide range of physical, chemical, and biological systems,\cite{wkunz04} often with significant impacts on bulk properties.\cite{petrache_salt_2006,zhang_interactions_2006} Various explanations have been put forward as to the origins of these fascinating phenomena.  Proposed contributing factors include: ion size,\cite{beggi08,vaitheeswaran_hydrophobic_2006} polarizability, \cite{jungwirth_ions_2002,dhagb05,tchan06} dispersion forces,\cite{wkunz04} and ion-water interactions.\cite{collins_ions_2007}  While there are known specific ion effects observed for cation sequences,\cite{collins_ions_2007} a great deal of recent attention has focused on anions due to their large size, polarizability, and dispersion 
interactions.\cite{zhang_interactions_2006}  As an example, 
Petrache {\it et al.}\cite{petrache_salt_2006} presented experimental results that show a large increase in the swelling of biological membrane multilayers when the Cl$^-$ ion is replaced with Br$^-$. The increase appears to be due to selective adsorption of Br$^-$ ions relative to Cl$^-$ ions at the membrane surface.  Other systems and phenomena impacted by specific ion effects\cite{wkunz04} include ion solvation free energies, ion activities, surface tension increments, bubble interactions,  colloid interactions, pH measurements, buffers, zeta potentials, ion channels,\cite{pusch_gating_1995} and protein stability.\cite{collins_ions_2007,zhang_interactions_2006,zhang_inverse_2009}

The segregation of certain ions to the liquid-vapor interface of water provides another example of ion specificity.\cite{vrbka_propensity_2004}  Experiments suggest that larger, often more polarizable, anions exhibit enhanced occupancy at the interface relative to bulk concentrations.\cite{mmuch05,raymond_probingmolecular_2004,petersen_nature_2006,ghosal_electron_2005,jchen06,jchen08}  These findings have implications for the understanding of atmospheric chemical reactions, sea-salt aerosols, and ions in biological systems. Molecular dynamics simulations have implicated the anion polarizability as a key determinant of surface specificity.\cite{mcari97,sagui05,jungwirth_specific_2006,mmuch05,tchan06}  The inclusion of anion polarizability for the halides appears to lead to an ion free energy minimum at water droplet or liquid-vapor interfaces in molecular dynamics simulations.\cite{lpere92,dhagb05,tchan06} Stuart and Berne,\cite{sstua96} utlizing a fluctuating-charge water model and a Drude oscillator polarizable ion model, came to a different conclusion: the larger dipoles in water due to induction seem more important for ion surface affinity than polarization effects on the chloride ion itself. As discussed above, other theoretical work has targeted dispersion interactions,\cite{wkunz04,bostrom_hofmeister_2005} ion-water interactions,\cite{lima_specific_2008,manciu_interactions_2005} and ion size.\cite{beggi08,vaitheeswaran_hydrophobic_2006}  There is clearly not a single agreed-upon mechanism that explains the observed experimental results. 

The above discussion begs the question of what level of theory is required to capture the essential physical effects leading to ion specificity.  If ion size and/or polarizability are the major determinants, then classical models with accurate representations of polarizabilities should be sufficient. The classical models assume, however, that fixed point multipole expansions and induced point dipoles accurately capture the quantum mechanical charge distributions in the condensed phase.  They also assume a pairwise form for the van der Waals dispersion interactions, thus neglecting possible many-body contributions. If complex charge rearrangements occur and/or there are appreciable many-body contributions to the dispersion energy, more accurate quantum mechanical treatments may be required. {\it Ab initio} molecular dynamics (AIMD) simulations at the gradient-corrected density functional theory (DFT) level are a common current choice for quantum modeling in the condensed phase. A well-known deficiency of DFT, however, is an improper handling of dispersion interactions. If those interactions play a signifcant role in ion specificity, a higher level of theory is required. As examples of the possible impact of a lack of proper dispersion interactions in DFT, the density of liquid water in AIMD modeling deviates from experiment by almost 20\%,\cite{mcgrath_simulating_2006} and cluster energetics are incorrectly estimated.\cite{santra_accuracy_2008} Addition of dispersion interactions at an approximate level improves the agreement of the DFT models with experiment.\cite{schmidt_isobaric-isothermal_2009}

An alternative approach is to measure or compute accurate frequency-dependent polarizabilities for the ions and bulk water, and then to utilize those polarizabilities in a Lifshitz approach for dispersion contributions to solvation free energies (or surface interaction free energies).\cite{parsons_ab_2009,parsons_nonelectrostatic_2009,barrydrew09}  From these quantum chemical studies, it is apparent that a wide range of imaginary frequencies is required to obtain converged interaction energies.  This strategy has sound fundamental underpinnings (as a many-body theory), but can fail to capture specific interactions of the ions with nearby waters in the condensed phase environment.  Another avenue developed recently is to compute free energies with a classical model and then correct the classical free energies via a perturbation formula;\cite{wood_free_1999,sakane_exploringab_2000,liu_hydration_2003} the quantum correction is estimated from averages of the Boltzmann factor for interaction energy differences between quantum and classical models calculated on configurations from the classical simulation.  Since quantum mechanics is not required to generate the trajectories, it is feasible to utilize correlated quantum mechanical methods (such as MP2-level theory). Mixed quantum mechanical/molecular mechanics (QM/MM) methods provide another viable approach for directly modeling ion solvation.\cite{ituno95,aohrn04} AIMD-DFT simulations coupled with thermodynamic integration methods have recently been employed to compute ion solvation free energies in water.\cite{leung_ab_2009}

In this paper, we take a step back from the difficult task of a high-level quantum treatment. As a first step, we examine a well-tested classical polarizable model for ions in bulk water. The purposes of this study are: 1) to illustrate the utility of quasi-chemical theory in disentangling physical components involved in ion hydration, 2) to determine the major factors driving ion specificity for hydration free energies in a classical polarizable model, and to test that classical model's accuracy, 3) to examine local solvation structure as functions of ion size and polarizability in the classical model, and 4) to develop methods and perform benchmark classical calculations which set the stage for future, more accurate QM/MM-based  calculations. By varying ion size and polarizability independently, we can gain insights into their relative contributions in ion hydration. This analysis is of course somewhat artificial physically, since the ion size and polarizability are typically related to each other. 

The model we choose for the present ion hydration studies is the AMOEBA force field.\cite{agros03,pren03,pren04} This model includes fixed partial charges, dipoles, and quadrupoles on each water atom, as well as induced dipoles on each atom.  The ions possess charges and induced dipoles.   The van der Waals (vdW) interactions are handled with a buffered 14-7 potential,\cite{thalg92} which has a replusive wall intermediate in steepness between the Lennard-Jones and Buckingham forms; this potential yields a relatively accurate estimate of dispersion interactions at the pairwise level.  The AMOEBA model produces a density maximum for pure water at 290 K and 1 atm.  In addition, the model accurately reproduces the experimental density, heat of vaporization, radial distribution functions, magnetic shielding, self-diffusion, and static dielectric constants of bulk water.  Also, ion parameters used with the AMOEBA water model have been shown to reproduce experimental hydration free energies.\cite{agros03} Alternative polarizable force fields exist, including the Dang-Chang model,\cite{tchan06} POL3,\cite{meng_molecular_1996} POL5/TZ,\cite{harder_polarizable_2005} NEMO,\cite{dhagb05} TTM2,\cite{gfano06} DPP,\cite{defusco_comparison_2007} Drude oscillator models,\cite{glamo06} and fluctuating charge models such as TIP4P/FQ.\cite{srick94,warren_hydration_2007,gwarr08}  Defusco, Schofield, and Jordan\cite{defusco_comparison_2007} have recently provided a careful comparison of several polarizable water models by analyzing cluster structure and energetics.  

The AMOEBA force field includes a Thole-type damping function that avoids a `polarization catastrophe' at short range.  Masia, Probst, and Rey\cite{mmasi06} showed that inclusion of a well-fitted damping function is crucial to model ion-water dipole moments properly. By comparison with AIMD-DFT simulations, Masia\cite{mmasi08} further showed that simply reducing the ion polarizability in the condensed phase is not sufficient to represent the distribution of instantaneous dipole magnitudes accurately. 
In that work, a longer-ranged Gaussian form for the damping
function was employed relative to the Thole version used in the AMOEBA force field, 
reflecting the diffuse electron densities of these ions.\cite{rashin_charge_2001} 
The Masia model accurately reproduces the distribution of ion dipole magnitudes from AIMD simulations. 
Finally, in a more extensive AIMD study, Guardia, Skarmoutsos, and Masia\cite{eguar09} observed slightly decreased water dipole moments near halide ions relative to bulk, and they compared the anion dipole distributions to a recent classical polarizable simulation (without damping) by Wick and Xantheas.\cite{cwick09}  The results show an over-polarization of the ions in the undamped simulation. Other classical polarizable models have employed reduced ion polarization models.\cite{ishiyama_molecular_2007} We will show below that the AMOEBA model also over-estimates the ion dipole magnitudes, even with the Thole damping function. This is likely due to the shorter range of the damping function relative to the model developed by Masia.\cite{mmasi08}

To address the aims stated above, here we utilize quasi-chemical theory to examine the influences of ion charge, size, and polarizability on ion hydration. We explore the relative contributions of first-order electrostatic, induction, and van der Waals interactions to the free energies, and also perform a structural analysis of the local solvation shell.
The remainder of the paper is structured as follows. In the next section, we briefly review quasi-chemical theory as applied to our analysis of ion hydration.  Then we discuss the computational methods employed in the present study.  The results of the simulations are presented, followed by a statement of our conclusions, discussion of the connections to related work, and future directions for investigation.  

\section{Quasi-chemical theory}

\begin{figure}
\vspace{-1.in}
\includegraphics[angle=0,width=1.\linewidth]{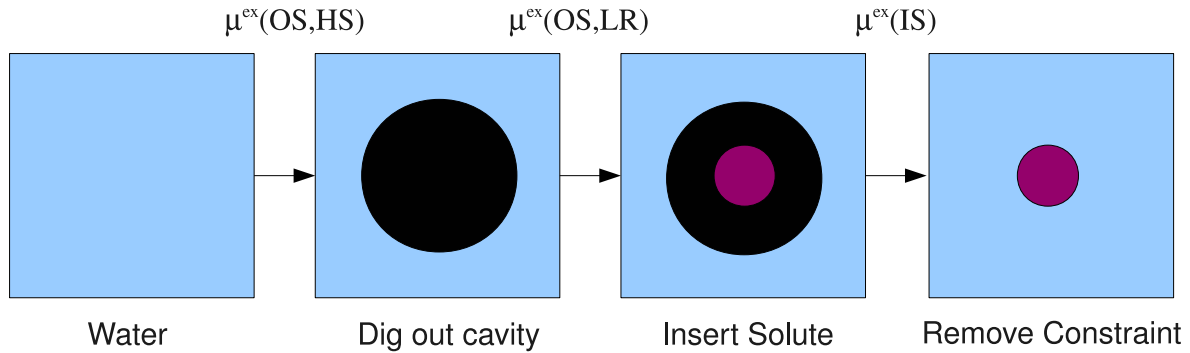}
\vspace{-2.2in}
\caption{Schematic representation of the quasi-chemical approach to solvation.}
\label{fig:qctschem}
\end{figure}

The quasi-chemical theory (QCT) breaks down the solvation free energy into a 
three-step 
process.\cite{ourbook,asthagiri_absolute_2003,asthagiri_quasi-chemical_2003,shah_balancing_2007,merchant_thermodynamically_2009,varma_structural_2008,droge08}  The steps include insertion of a hard sphere whose size is gauged by the length parameter $\lambda$, addition of the solute in the created cavity, and finally removal of the hard sphere to allow close contact of the solute and solvent (Figure \ref{fig:qctschem}).  Each term depends on $\lambda$, but the final result for the free energy is independent of $\lambda$.  

In the first step, a molecular-sized cavity is created in the solvent to accommodate a hard sphere solute.  The (outer-shell/hard-sphere, or packing) free energy required is given by
\begin{equation}
\label{eq:HS}
\beta \muex{OS,HS}(\lambda) = -\ln p_0(\lambda)
,
\end{equation}
where $p_0(\lambda)$ is the probability of spontaneous cavity formation (i.e. the solvent center closest to a defined point in solution has a distance further than $\lambda$).

Next, the solute is introduced into the cavity center.
This step can be considered in either the forward
(uncoupled,0$\rightarrow$coupled,1) or reverse (1$\rightarrow$0) directions, giving two equivalent formulas for the outer-shell, long-ranged free energy contribution:
\begin{align}
\label{eq:LR}
\beta \muex{OS,LR}(\lambda)
 &= -\ln \avg{0}{\efunc{-\beta \Delta U} | R_\text{min} > \lambda} \\
 &=  \ln \avg{1}{\efunc{\beta \Delta U} | R_\text{min} > \lambda} . \notag
\end{align}
For $\lambda$ values beyond a critical value $\lambda_{\mathrm{mf}}$, it is observed that the distribution of the interaction energies assumes a Gaussian form.\cite{shah_balancing_2007,droge08}
In this regime the contributions to $\Delta U$ come from many interacting
solvent molecules with similar interaction magnitudes.  This situation makes the
use of a quadratic perturbation formula a good approximation
to the true $\muex{OS,LR}$:
\begin{equation}
\label{eq:LRavg}
\beta \muex{OS,LR}(\lambda) \approx \left(
          \avg{0}{\beta \Delta U | R_\text{min} > \lambda} +
          \avg{1}{\beta \Delta U | R_\text{min} > \lambda}
  \right) / 2 .
\end{equation}
This equation can be obtained by performing a cumulant expansion of the two forms of Eq.~\ref{eq:LR} to second order and then averaging.  Since the distributions are assumed to be Gaussian, the fluctuation term is independent of the coupling state of the solute, and those terms cancel in the averaging. 

  An important point to notice about the Gaussian approximation to
$\muex{OS,LR}$ is that it implies a linear change in the average solute-solvent
interaction energy with respect to the solute coupling (traditionally
$\lambda$, but here we'll use $\gamma$):
\begin{align}
\avg{\gamma}{\Delta U | \lambda}
    &= \avg{0}{\Delta U | \lambda} - \beta \gamma \sigma^2 ,
\end{align}
with
\begin{align}
\sigma^2
    &= \avg{\gamma}{\delta{\Delta U}^2| \lambda}
     = \left(\avg{0}{\Delta U | \lambda} -
	\avg{\gamma}{\Delta U | \lambda} \right)/\beta\gamma .
\end{align}
The mean-field estimate of the long-ranged contribution is then
\begin{align}
\beta \muex{OS,LR}(\gamma;\lambda)
    &= \frac{\beta\gamma}{2} \left(
		\avg{0}{\Delta U | \lambda} +
		\avg{\gamma}{\Delta U | \lambda} \right) .
\end{align}
An example of a Gaussian approach is the Born model of solvation, where the coupling constant $\gamma$ scales the ion charge, and the Born radius is determined by the fluctuation contribution.  Below we will compare the electrostatic part of the long-ranged free energy contribution to the Born model prediction.

In the final step, the hard sphere condition is eliminated, allowing the
closest solvent molecules to form an inner solvation shell around the solute, yielding a chemical binding contribution to the free energy:
\begin{equation}
\label{eq:IS}
\beta \muex{IS} = \ln x_0(\lambda) ,
\end{equation}
where $x_0(\lambda)$ is the probability of spontaneous cavity formation around
the solute center with the solute included in the sampling. In practice, we use the Bayesian methods developed in Ref.~\cite{droge08} to compute $p_0 (\lambda)$ and $x_0 (\lambda)$.  In this approach, the observed solvent occupancies in shells around the cavity or solute center are employed to compute the free energies.   An alternative direct approach for computing $x_0$ and $p_0$ has been developed by Sabo {\it et al.}.\cite{sabo_studies_2008}

In a recent study, Merchant and Asthagiri \cite{merchant_thermodynamically_2009} provided an extensive classical analysis of ion solvation by carefully examining the thermodynamic contributions of particular coordination states to the inner-shell chemical term (Eq.~\ref{eq:IS}).  This analysis showed that the inner-shell term exhibits significant stepwise increments up to the most probable coordination number, with much less variation beyond that number.  The authors concluded that the ion influence on the solvent matrix is local based on these results. This important finding is consistent with the recent general discussion given by Collins, Neilson, and Enderby.\cite{collins_ions_2007}

In the present paper, we adopt a strategy that to some extent is counter to the original goals of quasi-chemical theory.\cite{ourbook,asthagiri_quasi-chemical_2003}  For the free energies computed here, we increase the value of $\lambda$ from 0 until a stable total solvation free energy is obtained, and term this value $\lambda_{\mathrm{mf}}$ (where `mf' refers to mean-field). Further increases of $\lambda$ do not alter the total free energy, indicating that by this value of $\lambda$ the distribution of interaction energies is nearly Gaussian.  We find that, for the spherical ion hydration problems examined here, a $\lambda_{\mathrm{mf}}$ value well inside the first maximum of the ion-water(oxygen) $g(r)$ is sufficient to yield Gaussian behavior for the long-ranged contribution.  Previous work has shown that non-Gaussian behavior is linked to high energy impacts of the waters with the solute, and a small level of conditioning removes the resulting high-energy tails.\cite{shah_balancing_2007,droge08} For the systems studied here, a good choice for $\lambda_{\mathrm{mf}}$ is the average distance of the closest water oxygen to the ion.  While we do examine the $\lambda$ dependence for a range of quantities, the above strategy was employed for most of the free energy calculations.  We also note that such a small value of $\lambda_{\mathrm{mf}}$ is not accurate for a molecular solvation problem such as the water molecule in water.\cite{droge08} For these anisotropic, molecular, hydrogen-bonded cases, the surrounding solvent must be pushed further out from the solute before Gaussian behavior is observed. 

With this choice for $\lambda_{\mathrm{mf}}$, the contribution of $\mu^{\mathrm{ex}}_{\mathrm{IS}} $ is very small (roughly -0.5 kcal/mol) for all three anions (Cl$^-$, Br$^-$, and I$^-$) and exhibits little ion specificity.  The packing contributions $\mu^{\mathrm{ex}}_{\mathrm{OS,HS}}$ are 5.3, 6.3, and 7.7 kcal/mol for the Cl$^-$, Br$^-$, and I$^-$ ions, respectively (below).  Those packing contributions depend only on the water model and $\lambda$.  So the  $\mu^{\mathrm{ex}}_{\mathrm{IS}} + \mu^{\mathrm{ex}}_{\mathrm{OS,HS}} $ combination for the chosen $\lambda_{\mathrm{mf}}$ values is positive and exhibits a small ion specificity mainly through the size-dependent packing term. The rest of the hydration free energy is then included in the outer-shell long-ranged contribution.  With this choice for the spatial division, the outer-shell long-ranged piece includes, besides interactions with distant waters, contributions from waters close to the ions.  

A physically based interpretation of quasi-chemical theory often includes the first solvation shell as the inner shell, handles that spatial domain with accurate quantum calculations, and treats the longer-ranged contributions either with a dielectric model or with classical simulations.\cite{asthagiri_quasi-chemical_2003} Here our aim is to start to disentangle the various energetic contributions to the ion hydration.  Pushing most of the problem into the long-ranged contribution that can be handled with a mean-field treatment allows for a division of the free energy into various parts: first-order electrostatic, induction, and vdW (defined below). There is no unique division of the free energy, but the present approach would appear to provide a natural and physical division.  A side product is that, once the inner-shell and packing contributions have been determined, the mean-field long-ranged contribution can be computed in relatively short simulation times since it simply involves energetic averages for the coupled and uncoupled cases.  This division should also prove useful in ongoing quantum mechanical studies of ion hydration.  

\section{ Computational Methods}
\label{sec:expt}

Molecular dynamics (MD) simulations were performed for several solutes in water using the NPT ensemble at 1 bar and 298.15 K. The Amber 10 package\cite{amber10} implementation of AMOEBA was employed for the simulations.  All simulations, involving 215 waters and the various solutes, employed velocity Verlet integration with a timestep of 1.25 fs, an induced dipole convergence parameter of $10^{-4}$ Debye, a tapered cutoff ending at 9 {\AA} for the vdW interactions, 
and their associated long-range corrections.
Each system was equilibrated for at least 60 ps using standard weak-coupling algorithms with temperature and pressure relaxation constants of 0.2 and 0.5 ps, respectively. This equilibration period is consistent with previous AMOEBA simulations of water and ions.\cite{pren03,agros03,laage_reorientional_2007} We observed no significant drift in any of the free energy components considered in the next section during block averaging of the data. The correlation times for the distance of closest approach for the closest water to the ion are less than 0.5 ps for all systems, and the average lifetimes for water occupancy in the first solvation shell are on the order of 10-20 ps.  In addition, we note that previous production runs were typically used as input for the next simulation. Since our simulations totalled over 20 ns in production runs, we are confident that the systems are properly equilibrated. Production runs of 1 ns duration used Andersen velocity randomization every 0.25 ps and a pressure relaxation constant of 1.5 ps during the collection of 2000 samples for analysis. Some additional simulations were performed using the nonpolarizable SPC/E water model for comparison.  


The anionic solutes studied were classified first by their vdW parameters as:
1, Cl$^-$-like, 2, Br$^-$-like, or 3, I$^-$-like, and second by their
polarizabilities ({\AA}$^3$): 0, $\alpha=0.0$ (nonpolarizable), 1, $\alpha=4.0$
(Cl$^-$-like), 2, $\alpha=5.65$ (Br$^-$-like) or 3, $\alpha=7.25$
(I$^-$-like).  All ion simulations employed the vdW parameters for the individual ions
from the AMOEBA code  
with no changes of well-depth or size.  Of course, the dispersion component of the vdW interactions depends on the ion polarizability; here, we omit this dependence and defer a careful analysis of the dispersion interactions to ongoing quantum mechanical studies. The ion size is then gauged by the the fixed set of vdW parameters; the distance at which the ion-water vdW potential crosses zero minus the radius of a water molecule gives a rough indication of that size for the three anions. 
Coupled solute-solvent simulations were run for each of these
solutes, and an additional coupled 500 ps sampling simulation including a Weeks-Chandler-Andersen (WCA)
model potential\cite{droge08} with radius around 3 {\AA} was run for systems corresponding to
the actual anions (1.1, 2.2, and 3.3).  For comparison to experiment a sodium
ion simulation (labeled by $+$) was also carried out.  Uncharged, nonpolarizable
solutes corresponding to the above particle sizes were also simulated and have
been denoted by the prefix $m$.  See Table \ref{tbl:models} for a compilation of the models
studied in this paper.  Properties of pure water were determined using
reference simulations including WCA particles with radii of $0$ (pure water),
$2.7$, $3.1$, and $3.4$~{\AA}.  In total, $20$ systems were simulated requiring
$21.2$ ns of MD, after the exclusion of system $1.3$ (Cl$^-$-like ion with $\alpha=7.25$~{\AA}$^3$), which exhibited sporadic numerical divergence during the induced dipole calculations.

\begin{table}[htbp]
  \begin{ruledtabular}
    \begin{tabular}{ *{4}{r}}
Model label& Charge & vdW & $\alpha$ \\
\hline
 +   &  +1 &  Na & Na  \\
\hline
 m.+ & 0 &  Na & 0  \\
 m.1 & 0 &  Cl & 0   \\
 m.2 & 0 &  Br & 0  \\
 m.3 & 0 &  I & 0  \\
\hline
 1.0 &  -1 &  Cl & 0    \\
 1.1 &  -1 & Cl & Cl \\
 1.2 &  -1 & Cl & Br \\
 2.0 &  -1 & Br & 0   \\
 2.1 &  -1 & Br & Cl \\
 2.2 &  -1 & Br & Br \\
 2.3 & -1 & Br & I \\
 3.0 &  -1 & I & 0  \\
 3.1 &  -1 & I & Cl \\
 3.2 &  -1 & I & Br \\
 3.3 & -1 & I & I 
    \end{tabular}
  \end{ruledtabular}
  \caption{Summary of the systems modeled in the present study. The labels are defined in the text. }
  \label{tbl:models}
\end{table}

  The Amber implementation of AMOEBA was validated against the Tinker program and found to give identical energy components when Amber's long-range Lennard-Jones correction was turned off.  Using Amber and systematically varying the Ewald parameter, $\eta$, between $0.4$ and 0.6 {\AA}$^{-1}$ for a chloride-water configuration carrying a net $-1$ charge revealed a drift in the electrostatic energy on the order of $0.28$ kcal/mol that could be removed by the addition of a uniform background charge correction term,
\begin{equation}
E_{\mathrm{c}} = -332.0716 \frac{Q^2 \pi} {2 V \eta^2} \text{  kcal/mol}
,
\end{equation}
where $V$ is the system volume and $Q$ is its net charge.  All energies reported here include this extra correction term (on the order of $-0.4$ to $-0.5$ kcal/mol depending on the system volume) not originally present in the raw Amber output.  A discussion of this net charge correction is given in Ref.~\cite{ourbook} (Eq.~5.50).

  For each system, the electrostatic potential at the center of the solute
(labeled $i$) due to the surrounding waters was estimated using numerical
differencing of $E_{\mathrm{es}}$ (the Amber electrostatic energy plus $E_{\mathrm{c}}$)  for charges of
$+/-10^{-2}$ electrons at zero solute polarizability.  This potential was
corrected for finite-size effects (Refs.~\cite{ghumm96} and \cite{ourbook}, Eq. 5.53) due to simulating with the
Ewald potential at a solute charge $q_\text{sim}$ via addition of $q_\text{sim}
\xi/L$, where $\xi = -332.0716 \times 2.837297$ kcal \AA\text{ }mol$^{-1}$ e$^{-1}$ for cubic simulation geometries of side length $L$:
\begin{align}
\label{eq:potl}
\Phi_i &= \left. \pd{E_{\mathrm{es}}}{q_i} \right|_{q_i=0} + q_\text{sim} \xi/L \\
  &\approx \frac{E_{\mathrm{es}}(q_i=10^{-2})-E_{\mathrm{es}}(q_i=-10^{-2})}{2 \times 10^{-2}}
     + q_\text{sim} \xi/L \notag
\end{align}

  Finally, the anisotropy of the solvation environment around a given solute
has been analyzed graphically by averaging the distance to the center of mass (COM)
of the nearest $n$ waters, generalizing a procedure used previously by other
authors.\cite{mcari97,sraug02}  First, all waters were rotated into a local
coordinate frame defined using the COMs of the nearest 3 waters.  This was done
in the usual way, using the first water COM to define the $x$ axis and
constraining the second to lie in the $x y$ plane with positive $y$ value.  We 
reflect about the $z$ axis to give the third
water's COM a positive $z$ value.  Next, the total COM of the nearest $n$
waters was averaged in this coordinate system.  Since it was found that the
major direction of change for this average with respect to $n$ was in the
$<1,1,1>$ direction, all COMs were projected along this direction to create
final one-dimensional anisotropy plots.  These plots show the tendency of water
to stack either in the direction of the first three solvent waters (positive values) or on the
opposite side (negative values).  We have found that the anisotropy plots illustrate the variation in solvation
structure more clearly than the unsigned distance distributions.

\section{Results}
\label{sec:results}

\subsection{ Cavity formation free energies}

  The AMOEBA water model has been parametrized to reproduce molecular multipole
moments, heats of vaporization, and structures of small water
clusters as well as bulk liquid density and internal energy.\cite{pren03,pren04}  It reproduces these quantities rather well, obtaining in addition good agreement with liquid phase radial distribution functions, heat capacity (21 +/- 5 cal/mol/K), average dipole moment in the liquid ($2.78$ D), and dielectric constant (82).  
For understanding ion hydration in the context of quasi-chemical theory, 
we have 
computed the cavity formation probability profile ($p_0$) as a function of $\lambda$, which yields the packing contribution to the free energy. 

\begin{figure}
\includegraphics[angle=-90,width=0.9\linewidth]{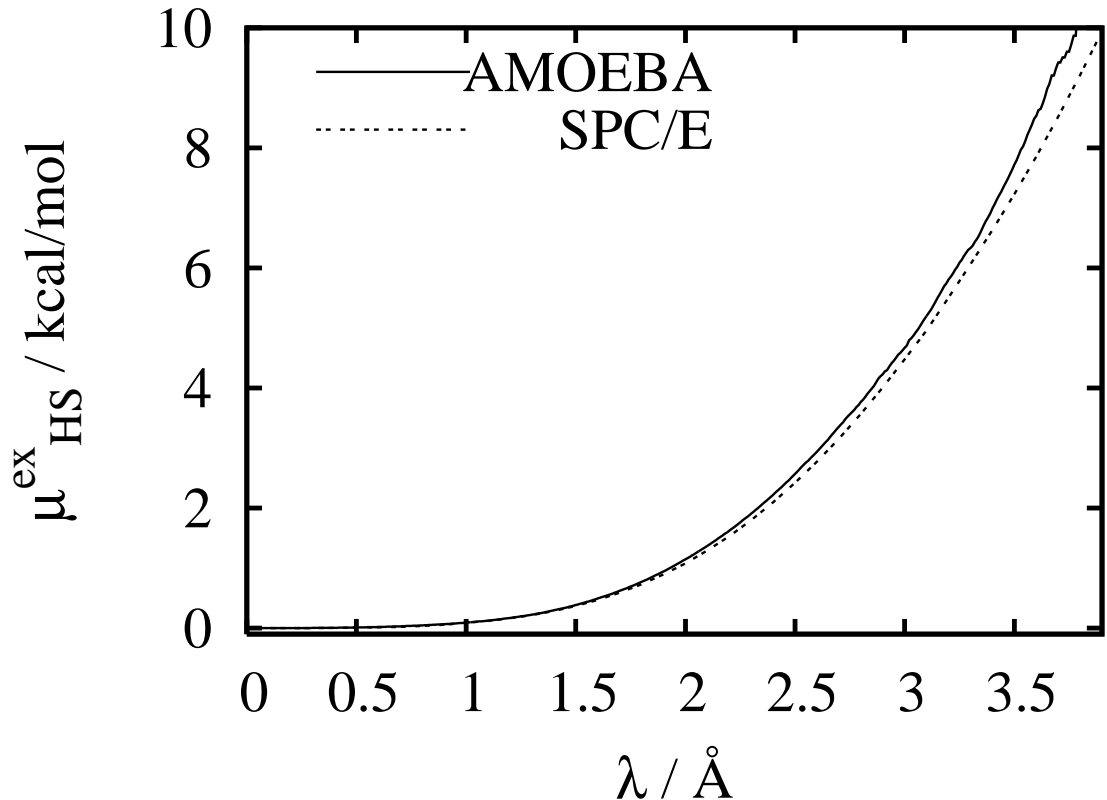}
\caption{Free energy cost for creating an oxygen unoccupied cavity of size
$\lambda$, $\muex{OS,HS}(\lambda)$, in NPT simulations of AMOEBA or SPC/E water models.}
\label{fig:p0}
\end{figure}

 Figure \ref{fig:p0} shows the hard sphere chemical potential profile derived
from the cavity occupancy probability for the AMOEBA and SPC/E water models.  
The major difference between the SPC/E and AMOEBA
water models for these profiles is an upward shift in the AMOEBA results that can be attributed to an
increase in the liquid phase density at 1 bar from $0.9556$ in SPC/E to
$1.0171$ g/mL in AMOEBA.  This AMOEBA value for the density is slightly larger than the
previous report of $1.0004$ g/mL,\cite{pren03} a difference which may be due
to the slightly longer time step, larger dipole convergence cutoff, or choice of temperature coupling algorithm for our calculations.  Despite this, the
difference in $\muex{HS}$ at $\lambda=3.5$ {\AA} is only $0.5$ kcal/mol.  From these results, it would appear that fixed-charge water models can provide accurate estimates of nonpolar solute hydration free energies.

\subsection{ Partial Molar Quantities}


  We have calculated the hydration free energies for all solutes studied in
this report and find excellent overall agreement with both previously
reported values and experimental whole salt free energies, as reported
in Table \ref{tbl:wholesalt}. The hydration free energies for the NaCl and NaBr pairs computed with the AMOEBA force field are within 0.2 kcal/mol of the recently tabulated experimental results in Refs.~\cite{mtiss98} and \cite{rschm00}. The AMOEBA result for the NaI pair is within 0.5-1.0 kcal/mol of the experimental values.  The error estimates were taken simply as the standard deviation over the several $\lambda$ values used to compute the free energies. As shown below in Figure \ref{fig:LRpart}, the errors come largely from the estimates of the (small magnitude) inner-shell and packing contributions, since the long-ranged part converges rapidly as a mean-field estimate. Thus it is likely the error estimates given are conservative and over-estimate the actual error in the total free energy calculations.



  The partial molar volumes of the AMOEBA ions computed
by differencing average simulation volumes agree moderately well with experiment within their stated uncertainties.  Despite the whole salt agreement, the single-ion partial molar volume data stand in contrast to Ref.~\cite{hfried73}, giving Na$^+$ a positive value at $5.0 +/- 2.3$ cm$^3$ mol$^{-1}$, and Cl$^-$ a correspondingly smaller value.  In the next section we show that the thermodynamic size decreases with increased polarizability, and note here that for system $1.2$ (Cl$^-$ size with $\alpha=5.65$~{\AA}$^3$), the computed partial molar volume becomes slightly negative.

\begin{table}[htbp]
  \begin{ruledtabular}
    \begin{tabular}{ *{5}{l}}
 $\muex{}$                      &  Na-Cl &  Na-Br &  Na-I  \\
 \hline
\mbox{Current}                  & -174.2 (1.6) & -167.7 (2.1) & -159.7 (2.0) \\
\mbox{Friedman\cite{hfried73}}  & -174.1       & -170.8       & -159.7 \\
\mbox{Tissandier\cite{mtiss98}} & -174.0       & -167.6       & -158.7 \\
\mbox{Schmid\cite{rschm00}}     & -174.0       & -167.6       & -159.2 \\
\hline
 $\partial V / \partial N$      &  &  &  \\
 \hline
\mbox{Current}                  &  20.5  (3.3) &   22.5 (3.4) &   41.1 (3.2) \\
\mbox{Friedman}                 &  16.6        &   23.5       &   35.0 \\
\mbox{Krumgalz\cite{bkrum96}}   &  16.62       &   23.479     &   34.998 
    \end{tabular}
  \end{ruledtabular}
  \caption{Partial molar hydration quantities for whole salts at infinite
dilution, in kcal/mol and cm$^3$ mol $^{-1}$. Numbers in parentheses are
one standard deviation errors. }
  \label{tbl:wholesalt}
\end{table}

  Because it is experimentally impossible to measure single-ion partial molar
quantities, proton hydration is usually employed as a reference process.  The
present simulations can be used to infer values for this process.
The infinite dilution free energies and enthalpies reported here, however,
neglect issues associated with the average potential of the liquid phase
because they lack a liquid/vapor interface.  Energies calculated using this
surfaceless approach have been termed `intrinsic' and can be corrected for the
presence of a surface potential by adding $q \phi_{\mathrm{vl}}$.

  The proton solvation
quantities were predicted by combining our data with experimental H$^+$-Na$^+$ differences as well as direct measurements of halide acids.  Intrinsic free energies calculated in this way vary little over estimates from alternative ions, giving $-252 + \phi_{\mathrm{vl}}$ kcal/mol with a simulation error of less than $1$ kcal/mol, in relatively good agreement with earlier work.\cite{topol_structure_1999,agros03,asthagiri_absolute_2003,glamo06}

  The surface potential also creates a possible ambiguity in the partial molar volume, $\bar V = \pd{\mu_\text{real}}{p} = \pd{\mu}{p} + q \pd{\phi_{\mathrm{vl}}}{p}$, emphasizing the dependence of the surface properties on the experimental conditions.  The data on
partial molar volumes for HCl and sodium salts have been used to calculate
an intrinsic proton partial molar volume.  The computed estimates made using these
different experimental values, however, vary between $0.1$ and $6.2$ cm$^3$ mol$^{-1}$ with our simulation uncertainty on the order of $2.3$ cm$^3$ mol$^{-1}$.


\subsection{ Free Energy Components}

\begin{figure}
\includegraphics[angle=-90,width=0.8\linewidth]{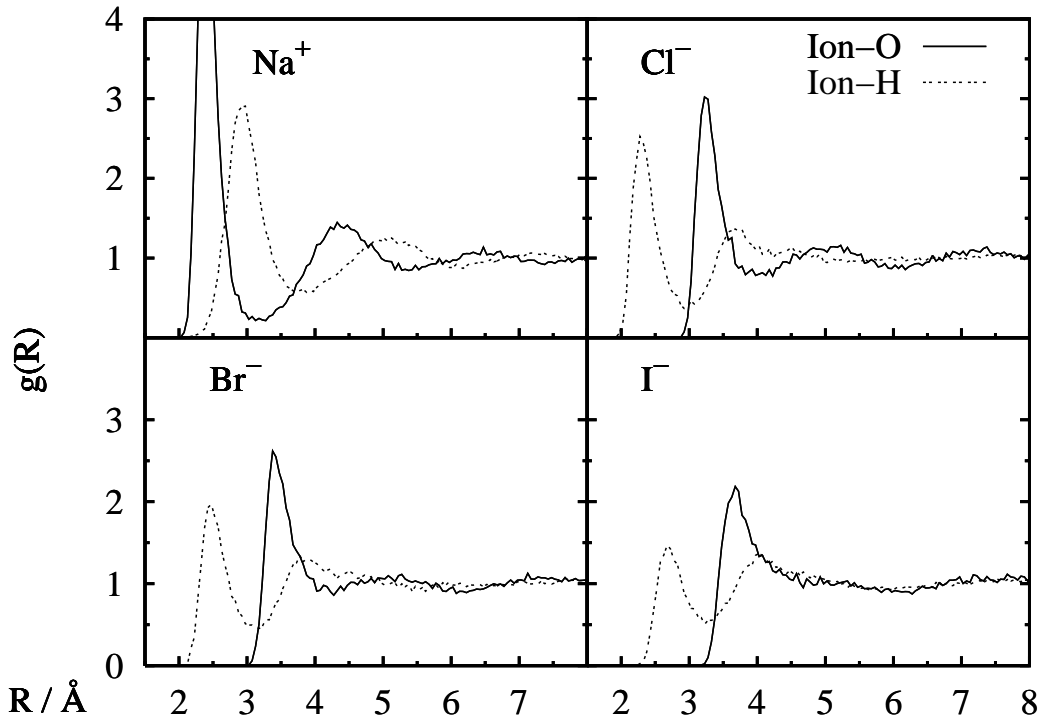}
\caption{Ion-water(hydrogen) and ion-water(oxygen) radial distribution functions for the Na$^+$, Cl$^-$, Br$^-$, and I$^-$ ions modeled with the AMOEBA force field.}
\label{fig:rdfs}
\end{figure}
\begin{figure}
\includegraphics[angle=-90,width=0.9\linewidth]{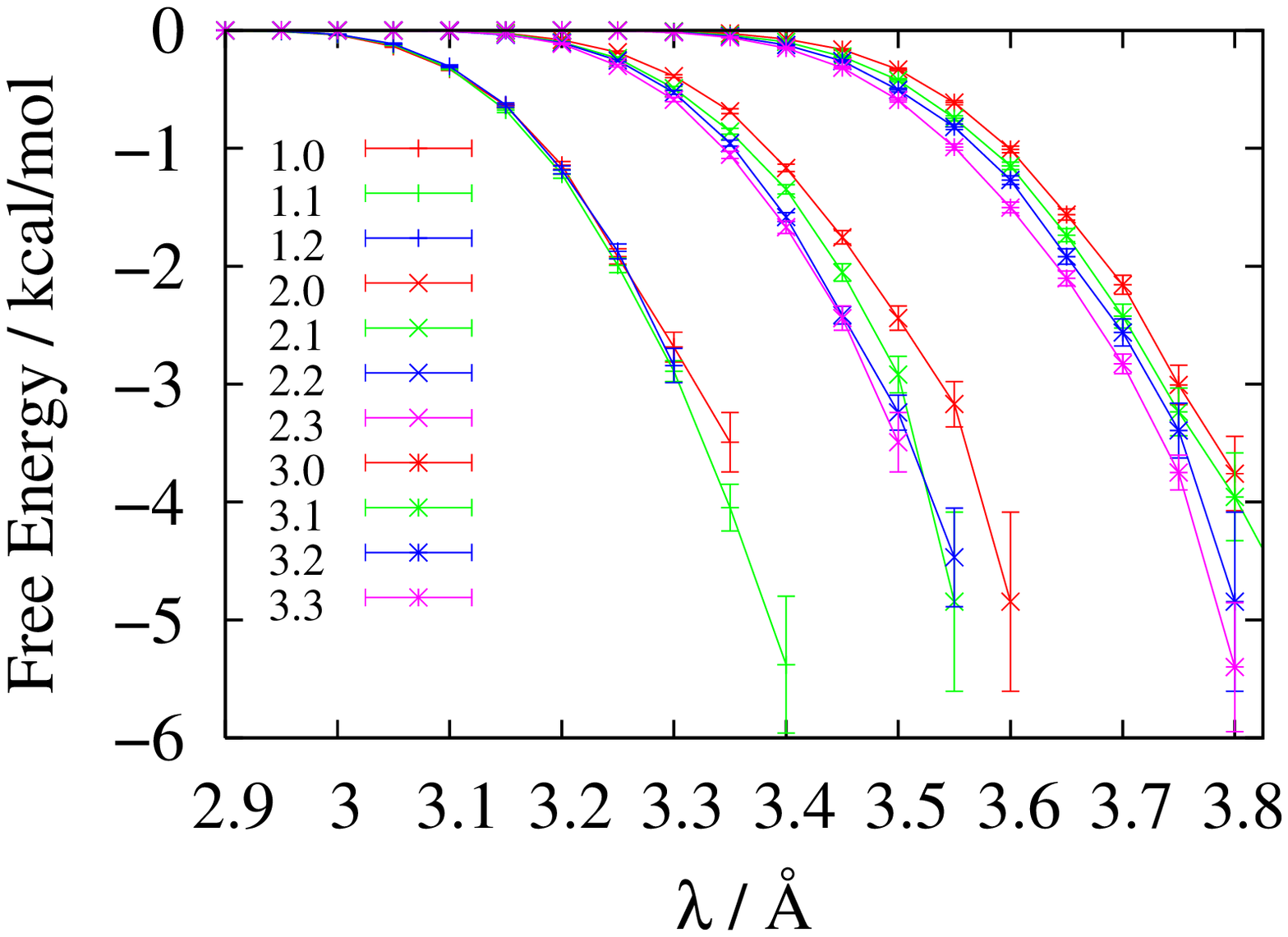}
\caption{Effect of polarizability on the energy of collapsing the inner-shell partition. For ions of all sizes (first index: 1=Cl,2=Br,3=I), the curve shifts downward with increasing polarizability (second index: 0=zero,1=Cl,2=Br,3=I) indicating that increased polarization causes water to pack more closely.}
\label{fig:x0}
\end{figure}

As discussed above,  quasi-chemical partitioning has 
been used to separate the total hydration free
energy into physical components. The ion-water radial distribution functions (rdfs) are shown in Figure \ref{fig:rdfs} for reference.  The $\lambda$-dependences of the chemical contribution from inner-shell
waters for each anion studied are shown in Figure \ref{fig:x0}.  At first, it
may be surprising that there is no identifiable jump due to breaking of
direct-contact interactions with water which would be expected around $3.1$--$3.3$, 
$3.3$--$3.45$, or $3.5-3.7$ {\AA} for Cl$^-$, Br$^-$, or I$^-$, 
respectively.\cite{srame08} The above distances correspond to locations just beyond the first minimum in the ion-water(hydrogen) rdf, or roughly at the first maximum of the 
ion-water(oxygen) rdf.
Instead, the data show a smoothly increasing energetic penalty for pushing
waters away from the anion -- indicating the role of more distant waters (perhaps through many-body polarization effects) in
creating a positive potential at the ion's center that compensates for the disruption of direct contacts.

  Increasing the ion 
polarizability results in a systematic downward shift
in the $\muex{IS}(\lambda)$ profile, reflecting an increased probability of inner-shell
water occupancy.  This provides direct evidence for the decrease in
thermodynamic ion size with increasing polarizability previously inferred by
several authors.\cite{lpere92,mcari97,dhagb05,gwarr08}  This finding is also
reflected by the trend in partial molar volumes discussed above.  It is well known that the introduction of charge on a neutral solute has an even  larger impact on the local solvation structure.

The $\mu^{\mathrm{ex}}_{\mathrm{IS}} (\lambda)$ profiles presented in Figure \ref{fig:x0} are computed out to $\lambda$ values that are just beyond the location of the first maximum in the ion-water(oxygen) rdfs.  For the total free energy calculations discussed below, the $\lambda_{\mathrm{mf}}$ values used were taken to be the average distances to the closest water oxygen, $\langle R_{\mathrm{min}} \rangle_1$, which are roughly 3.12, 3.29, and 3.5 {\AA} for the Cl$^-$, Br$^-$, and I$^-$ ions, respectively.  The magnitudes of $\mu^{\mathrm{ex}}_{\mathrm{IS}}$ for these values of $\lambda$ are quite small (on the order of -0.5 kcal/mol or less), and exhibit little polarizability dependence.  

\begin{figure}
\includegraphics[angle=-90,width=0.9\linewidth]{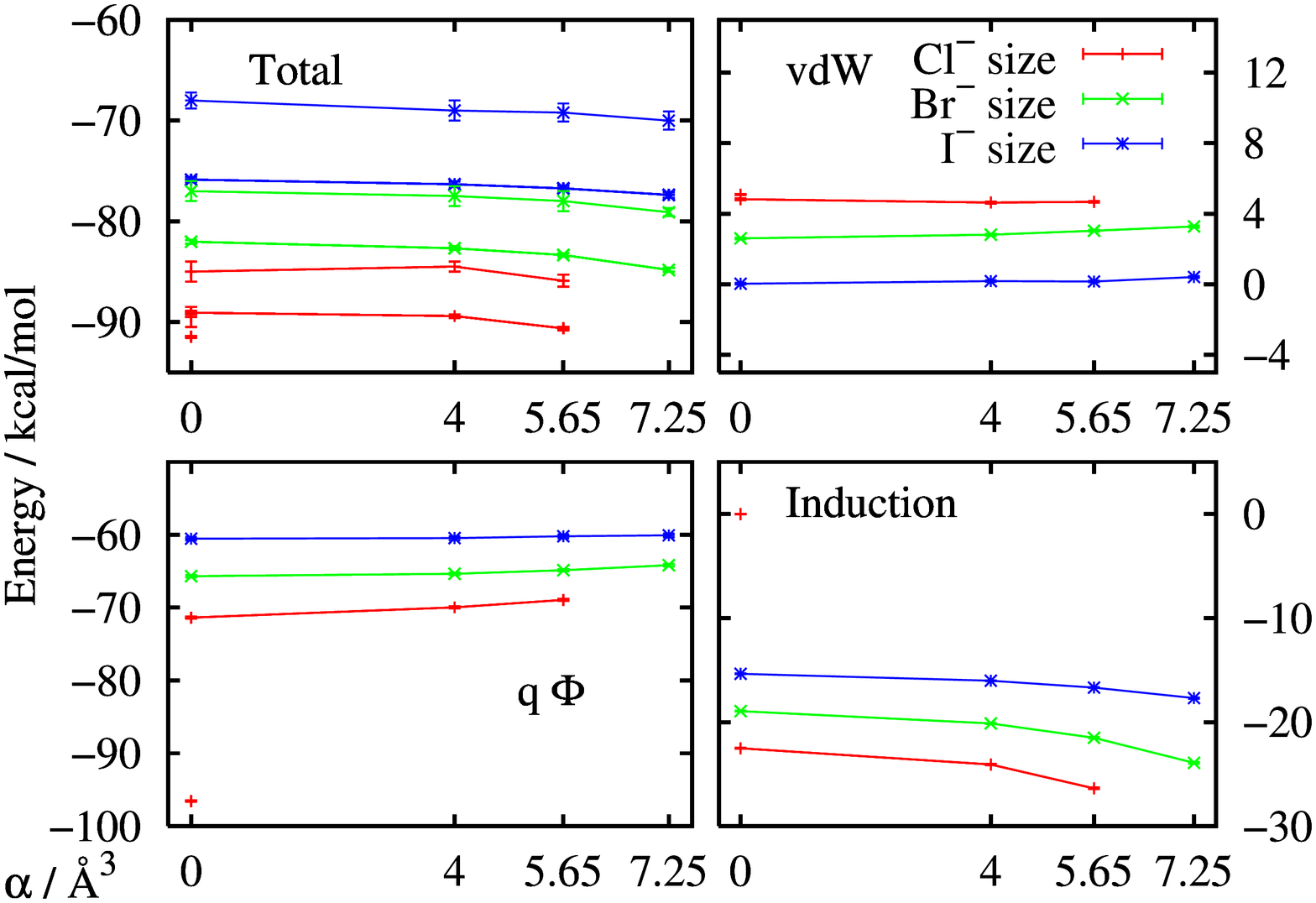}
\caption{Contributions to the OS-LR conditional free energy average at 
$\lambda \approx \avg{1}{R_\text{min}}$.  As explained in the text, the components are averages over the coupled and uncoupled cases.  The corresponding contributions from a nonpolarizable Cl$^-$-SPC/E system are shown as single points. In the upper left-hand figure, the upper curves for each ion are the total hydration free energies obtained by adding the inner-shell and outer-shell packing contributions.}
\label{fig:LRpart}
\end{figure}

Because the inner-shell contribution is so small for the chosen $\lambda_{\mathrm{mf}}$ (and shows little ion polarizability dependence), and the packing term is a well-characterized size-dependent contribution independent of ion polarizability,
 the origin of the observed free energy
differences between ions can be investigated by 
examination of the outer shell, long-ranged
free energy component (Figure \ref{fig:LRpart}).  As described above, for $\lambda$ values large enough to yield mean-field behavior, $\muex{OS,LR}$ is the average of
the mean interaction energy between the coupled and uncoupled simulations.
This allows for a simple, physically 
motivated division of the free energy for this step in terms
of the first-order electrostatic (es), induction (ind), and van der Waals (vdW) energies.


  The division of $\Delta U$ is the usual one for force field simulations,
with the exception that the electrostatic energy difference, $\Delta E_{\mathrm{es}} =
E_{\mathrm{es}}(\text{solvent + solute}) - E_{\mathrm{es}}(\text{solvent})$, includes many-body induction
due to the instantaneous changes in the system dipoles in the presence of the
solute charge, $q_i$. We define $\Delta E_{\mathrm{es}}$ as
\begin{align}
\Delta E_{\mathrm{es}} \equiv q_i \Phi_i + \Delta E_{\mathrm{ind}} .
\end{align}
The potential at the solute charge location due to the surrounding solvent's
self-consistent dipole configuration (but uncoupled from the solute), $\Phi_i$, is as defined in Eq.~\ref{eq:potl}. The $\Delta E_{\mathrm{ind}}$ term includes the remainder of the electrostatic energy change, which is the energy change to repolarize the ion and the waters as they mutually interact.  We note that a size-dependent correction ($\xi q_i^2/2L$, where $\xi = -2.837297$ and $L$ is the box size) is omitted from $\Delta E_{\mathrm{es}}$ here since it has no impact on the computed free energies. 

  For nonpolarizable force fields, the induction energy is zero.
For polarizable force fields, the potential/induction energy separation can be
likened to the symmetry-adapted perturbation theory (SAPT)\cite{stone,amisq05,bukowski_predictions_2007}
partitioning of interaction energies employed in quantum calculations. Along these lines, we term the $q_i \Phi_i$ piece the `first-order electrostatic energy' and the remainder will be called the induction contribution.  
The remaining contribution $\Delta E_{\mathrm{vdW}} = E_{\mathrm{vdW}}(\text{solvent +
solute}) - E_{\mathrm{vdW}}(\text{solvent})$ has the traditional interpretation
of repulsion plus pairwise dispersion energy.
These three parts, then, make up the total solute-solvent interaction energy
for any given configuration of the classical model:
\begin{equation}
\Delta U = q_i \Phi_i + \Delta E_{\mathrm{ind}} + \Delta E_{\mathrm{vdW}} .
\end{equation}
Plots of the estimated free-energy contributions obtained from this partitioning as functions of the solute size and
polarizability, $\alpha$, are presented in Figure \ref{fig:LRpart}.   The energy 
components have been calculated at a $\lambda$ corresponding to the average
distance to the nearest water oxygen $\avg{1}{R_\text{min}}$ as listed above.  The free energies calculated
from these radii are in good agreement with the estimates over all $\lambda$ used
in creating Table \ref{tbl:wholesalt}.

As can be seen in the upper-left panel of Figure \ref{fig:LRpart}, the total free energy and total long-ranged contribution show a much larger dependence on ion size than on polarizability, consistent with free energy calculations using other polarizable models.\cite{warren_hydration_2007} The addition of the $\mu^{\mathrm{ex}}_{\mathrm{IS}} + \mu^{\mathrm{ex}}_{\mathrm{OS,HS}}$ combination does not alter the polarizability dependence but does exhibit some specificity through a purely size-dependent effect in $\mu^{\mathrm{ex}}_{\mathrm{OS,HS}}$ (above), further separating the hydration free energy differences.  The first-order electrostatic contribution is the largest part of the long-ranged contribution, and it exhibits size-dependent specificity and a slight decrease in favorability with increasing polarizability.  As expected, the induction contribution displays the largest dependence on polarizability, and the observed downward trend with increasing polarizability overcomes the reverse trend in the first-order electrostatic part.  The vdW piece exhibits an ordering that is inverted relative to the electrostatic parts, with the Cl$^-$ ion displaying the largest positive value. Thus this vdW part of the long-ranged contribution serves to counteract somewhat the size-driven separation in free energies due to the electrostatic terms. It appears that the attractive dispersion component for the larger Br$^-$ and I$^-$ ions counteracts the more repulsive contribution for the Cl$^-$ ion.  Interestingly, the combination of the outer-shell packing 
and the vdW part of the outer-shell, long-ranged contributions ($\mu^{\mathrm{ex}}_{\mathrm{OS,HS}} + \mu^{\mathrm{ex}}_{\mathrm{OS,LR}} (\mathrm{vdW})$) shows little specificity for the Cl$^-$,Br$^-$, I$^-$ sequence: 11.2, 11.1, and 11.9 kcal/mol, respectively. Since the $\mu^{\mathrm{ex}}_{\mathrm{IS}}$ also displays little specificity for the chosen $\lambda$ values,  these results suggest that most of the specificity is carried in the electrostatic contributions to $\mu^{\mathrm{ex}}_{\mathrm{OS,LR}}$ in the classical model. 

\begin{figure}
\includegraphics[angle=-90,width=0.9\linewidth]{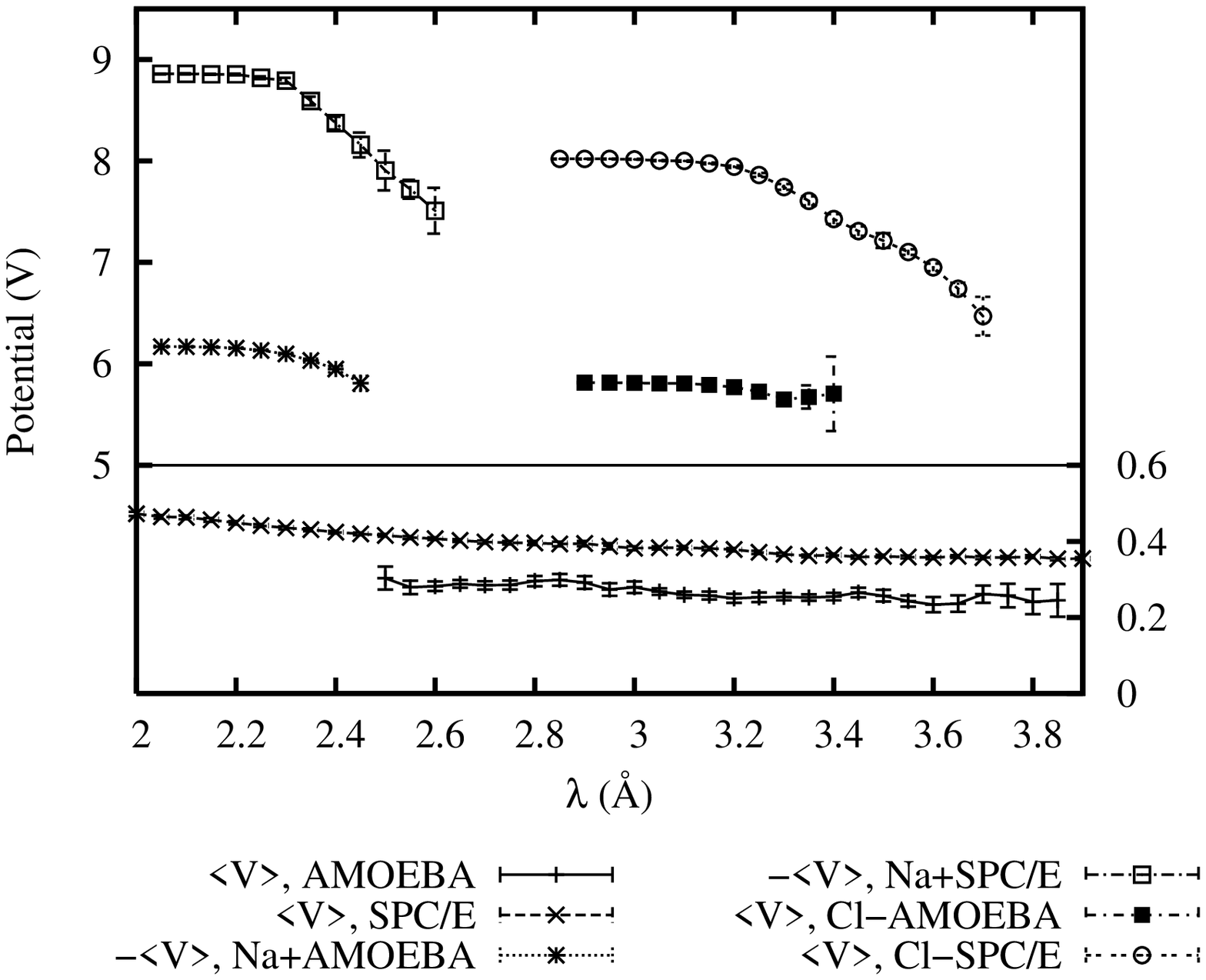}
\caption{Electrostatic potential at the center of a test solute in NPT
simulations of AMOEBA or SPC/E water models: (lower section) hard spheres of
size $\lambda$, (upper section) Na$^+$ and Cl$^-$ ions (left and right of plot, resp.).}
\label{fig:potl}
\end{figure}

In order to better understand the relative roles of the various parts of the electrostatic interactions, we here examine the electrostatic potentials $\Phi_i$ at cavity and ion centers  under various circumstances.  
The orientation of water molecules in the vicinity of a neutral
cavity gives rise to a positive electrostatic potential at its center.\cite{ghumm96,agros05}  This
potential is plotted as a function of $\lambda$, the distance to the nearest
water oxygen, in the lower section of Figure \ref{fig:potl}.  The AMOEBA potential is slightly higher ($\approx 60$ mV) than that at the center of an uncharged
solute\cite{agros05} because our cavity definition allows intrusion of water
hydrogens, which would otherwise be repelled.  Although both AMOEBA and 
SPC/E water models share
a similar trend, the charge distributions in the AMOEBA model serve to shift
the cavity potential downwards by $\approx 100$ mV.  This is interesting because, while the AMOEBA fixed partial charges are reduced in magnitude relative to the SPC/E model, the average dipole magnitudes for the AMOEBA waters near the cavity (2.76 D) are close to the AMOEBA bulk value (2.78 D) and are larger in magnitude than the fixed permanent dipole of the SPC/E model (2.35 D). These AMOEBA dipole magnitudes near the cavity are
in sharp contrast to the marked decrease in the dipole moments approaching the
water/vacuum interface.\cite{tchan06}  We can conclude that many-body polarization effects are important in orienting the dipoles around small cavities. 

The first-order electrostatic contribution to the long-ranged part of the free energy comes from the average of the potential for the uncoupled and coupled cases. 
To examine the coupled case, we have added to Figure \ref{fig:potl} the
electrostatic potential at the center of ionic solutes (Na$^+$ and Cl$^-$)  
as a function of the
distance to the nearest water oxygen.  The ion models used with the SPC/E water model were the
same as in our previous publication,\cite{droge08} while the AMOEBA ions are
taken directly from the present study.  First, the first-order potential $\Phi_i$ for the coupled case is much larger for the SPC/E model than for the AMOEBA model, reaching values close to -9 V for the Na$^+$ ion at small $\lambda$ values. Second, there is a clear break in slope for the SPC/E model at $\lambda$ values close to the first maximum in the ion-water(oxygen) rdf; pushing the waters further away from the favorable solvation environment creates a large energetic penalty.  The polarizable AMOEBA model, on the other hand, softens this effect through many-body polarization effects.  For the SPC/E case, the total electrostatic part of the long-ranged free energy contribution comes entirely from the first-order electrostatic contribution. The AMOEBA model makes up the remainder of the free energy through induction effects, and these are seen to be substantial in magnitude (about 25\% of the total electrostatic part of the free energy). 
Making the above plot with a nonpolarizable Cl$^-$ ion in AMOEBA
does not significantly change the shape of the potential profile, but shifts it
upward by $\approx$ 140 mV.  We conclude that many-body polarization contributions are significant in representing the electrostatic potential in nonpolar cavities and ion centers.  The cavity free energy is rather insensitive to polarization, however, as seen in Figure \ref{fig:p0}.   

\begin{figure}
\centering
\includegraphics[angle=-90,width=0.9\linewidth]{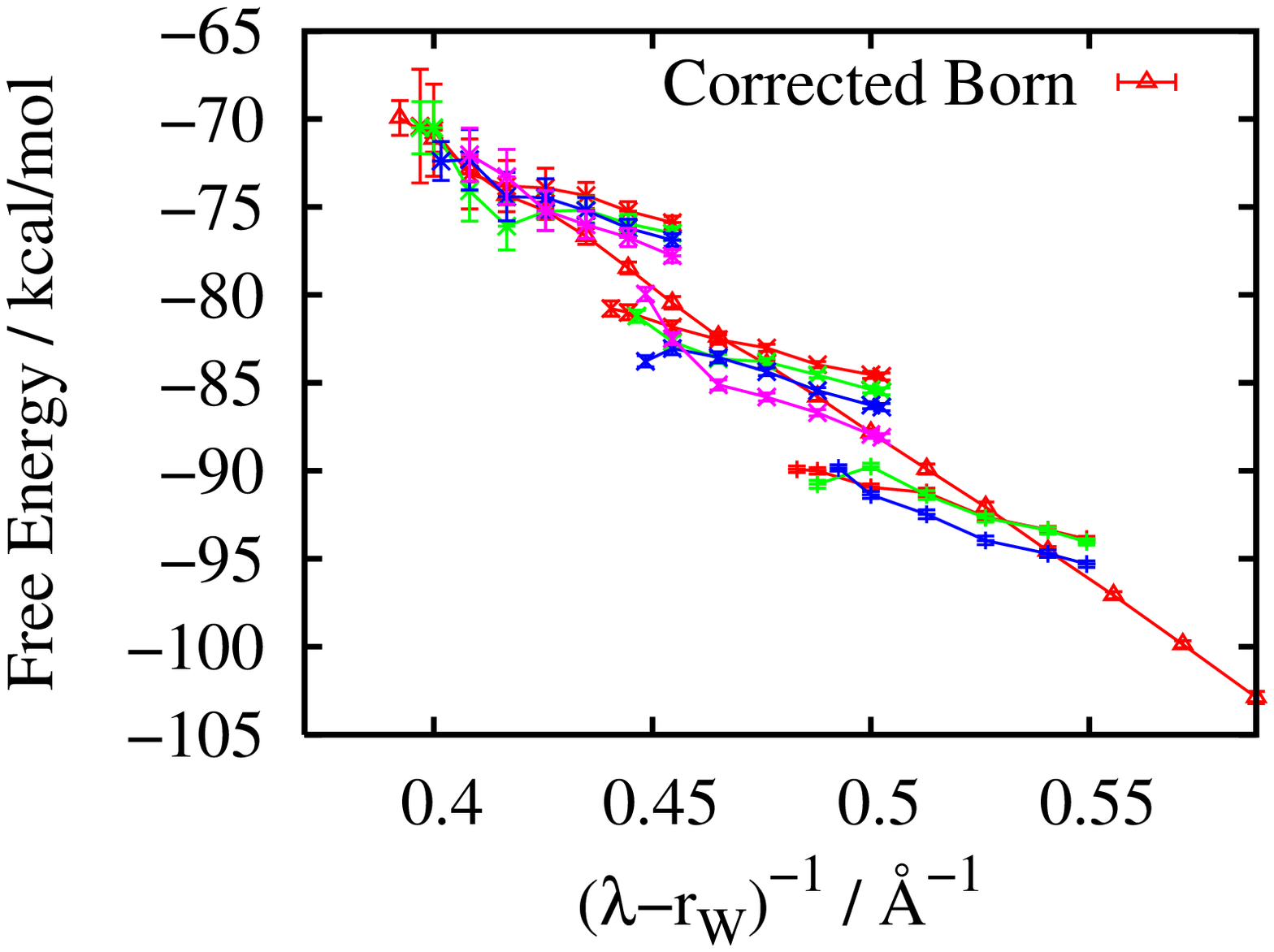}
\caption{Effect of size and polarizability on the total electrostatic
contribution to the OS-LR free energy component.
Decreases in size and, to a lesser extent, increases in 
polarizability lead to more favorable interactions.  Interestingly, variations in
the ion sizes produce about a 5 kcal/mol separation when compared at equal
nominal radii, contradicting the Born model prediction. Labels are as in Figure \ref{fig:x0}.}
\label{fig:born}
\end{figure}

As a point of interest, the complete electrostatic contribution to the long-ranged part of the free energy can be compared to the Born model, $\muex{OS,LR,es} \approx
 q_i \Phi_0 (\lambda) - (1-1/\epsilon)\tfrac{q_i^2}{2 R}$ (corrected for nonzero cavity potential, $\Phi_0(\lambda)$,
Figure \ref{fig:potl}); $R$ is $\lambda - R_W$, where $R_W$ is the radius of the water molecule, taken here as 1.3 {\AA}.  When such a comparison is made (Figure \ref{fig:born}), 
we see that in a general sense the Born model is relatively accurate even at these short length scales near the ion.  Since the Born model is equivalent to the Gaussian model employed here for the OS,LR contribution,\cite{ghumm96} the agreement is not surprising.  On the other hand, it is clear that molecular details matter in obtaining an accurate representation of the electrostatic part of the free energy,\cite{hashb00,pyang08,skalko96} since the slopes of the AMOEBA-generated curves differ from the Born model, and there are noticeable jumps between ions for a given $\lambda$ value.  

\subsection{Analysis of local solvation structure}

\begin{figure}
\includegraphics[angle=-90,width=0.8\linewidth]{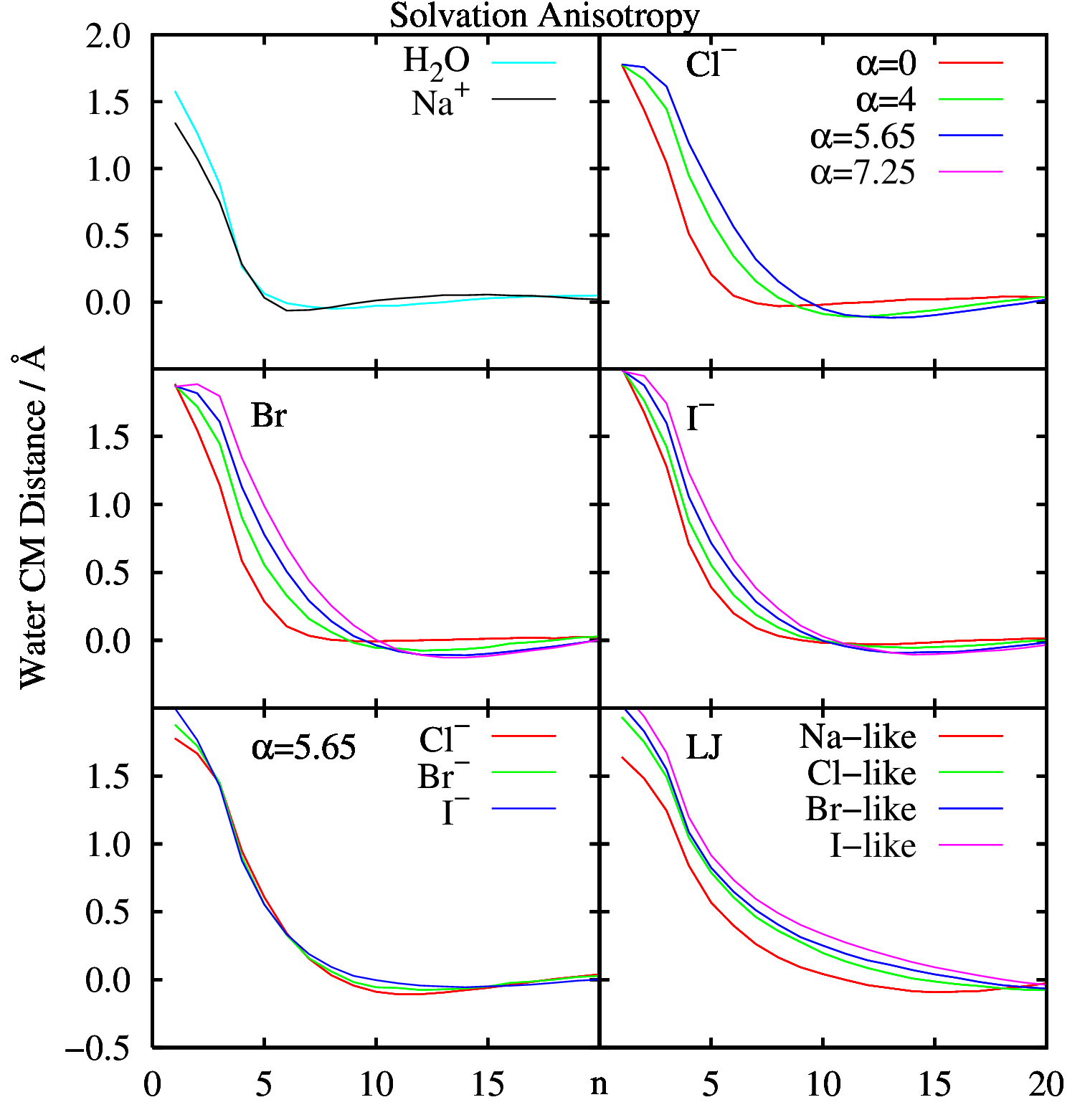}
\vspace{4.5in}
\caption{Effect of size and polarizability on the solvation shell organization.
Plotting the projection of the average center of mass of the closest n solvent waters onto a vector in the direction of the first three shows how quickly subsequent layers of solvent relax to an isotropic distribution. Labels are by polarizability parameter with the exception of the lower right panel which shows uncharged, nonpolarizable solutes with the same vdW parameters as the corresponding ions.}
\label{fig:aniso}
\end{figure}

In addition to the thermodynamic and electrostatic analysis above, it is interesting to explore the local solvation structure around the various ions as functions of size and polarizability.  
In Figure \ref{fig:aniso}, the solvation-shell water ordering is addressed by plotting
the signed center of mass distance of solvating water clusters of 
increasing size (above), ordering waters using the
distance from the ion to their center of mass.  A recent
report\cite{cwick09} examined the same question by projecting along the
direction of the instantaneous ion dipole.  Yet another possibility would be to search directly for molecular cavities in the solvation water by, for example, computing a potential of mean force between the ion and a hard sphere.  

  This anisotropy is significant because it shows one of the mechanisms by which
ordering may increase in response to increased polarization.  It also indicates a
possible role of polarization on surface affinity, since a rightward shift of
the first zero in the anisotropy plot may indicate a higher solute preference
for surface-like solvation shells.\cite{cwick09}  

We measure the degree of solvation asymmetry by the number of waters just before the anisotropy plot reaches its first zero.  Table \ref{tbl:comp} presents the numerical values for these cluster sizes, along with a range of other solvation data. 
Figure \ref{fig:aniso} displays the solvation anisotropy for a range of cases.  In the upper-left panel, it is apparent that the water molecule and the Na$^+$ ion exhibit relatively symmetric solvation shells in comparison with the three anions and nonpolar solutes. For the three anions, it is clear that increasing the ion polarizability shifts the curves rightward, indicating increased anisotropy.  The anisotropy also increases through the sequence Cl$^-$, Br$^-$, I$^-$, for their normal AMOEBA parameter values (systems 1.1, 2.2, and 3.3). In the lower-left panel, we present the anisotropy plot for the sequence Cl$^-$, Br$^-$, I$^-$ but with a fixed ion polarizability of $\alpha = 5.65$~{\AA}$^3$.  Surprisingly, there appears to be little size dependence to the anisotropy.  The final lower-right panel displays the anisotropy for uncharged solutes with vdW parameters corresponding to each ion.  For uncharged solutes, there is a clear size dependence, and the water solvation shell is quite asymmetric. 

\begin{table}[htbp]
  \begin{ruledtabular}
    \begin{tabular}{ *{6}{r}}
System& $N_0$ & $<R_\text{min}>_1$ & $\avg{}{D_{\mathrm{I}}}$ & $\avg{}{D_{\mathrm{W}}} $ & \muex{} 
  \\
\hline
 +   &   5&  2.279(2) &0.0238(3) & & -87.7(11) \\
\hline
 m.+ & 10 &  2.837(8) & 0.0  &   & 1.5(2) \\
 m.1 & 14 &  3.344(6) & 0.0  &   & 2.8(3) \\
 m.2 & 16 &  3.492(5) & 0.0  &   & 3.1(3) \\
 m.3 & 18 &  3.681(7) & 0.0  &   & 3.5(5) \\
\hline
 1.0 &  6 &  3.116(2) & 0.0    & 2.57(06) & -85.0 (10) \\
 1.1 &  8 & 3.115(2) & 1.56(2)& 2.72(05) & -84.5 (5) \\
 1.2 &  9 & 3.119(2) & 2.92(4)& 2.61(06) & -85.9 (6) \\
 2.0 &  8 & 3.304(3) & 0.0    & 2.67(05) & -77.0 (10) \\
 2.1 &  9 & 3.289(3) & 1.16(2)& 2.68(02) & -77.5 (10) \\
 2.2 &  9 & 3.281(3) & 2.19(3)& 2.69(09) & -78.0 (10) \\
 2.3 & 10 & 3.274(2) & 3.61(5)& 2.66(04) & -79.1 (4) \\
 3.0 &  9 & 3.515(3) & 0.0    & 2.70(07) & -68.0 (8) \\
 3.1 &  9 & 3.496(3) & 0.80(1) & 2.63(04) & -69.0 (10) \\
 3.2 &  9 & 3.485(3) & 1.48(3)& 2.61(07) & -69.2 (9) \\
 3.3 & 10 & 3.470(3) & 2.37(4)& 2.65(07) & -70.0 (9) 
    \end{tabular}
  \end{ruledtabular}
  \caption{Comparison of local solvation environment indicators: $N_0$, the
number of waters required for Figure \ref{fig:aniso} to reach zero, average
distance to the closest water oxygen (\AA), average ion dipole moment
(Debye), average first-shell water dipole moment (Debye), and single-ion total hydration free energy (kcal/mol).  Numbers in parentheses indicate numerical uncertainty in the last digit. The label $m$ indicates an uncharged solute. }
  \label{tbl:comp}
\end{table}

Table \ref{tbl:comp} collects the solvation environment properties relevant
to this discussion. The total hydration free energies are also presented for each case for reference.  The number of waters just before the anisotropy plot first crosses zero, $N_0$, is largest for the uncharged, non-polarizable solutes indicating the highest asymmetry for those cases.  The Na$^+$ ion exhibits the least asymmetry, and $N_0$ appears to correlate most strongly with polarizability for the anion series.  The most probable solvation number (the number of waters with hydrogens within 3.8~{\AA} of the ion and with an ion-H-O angle greater than $130^{\circ}$) is 6 for all of the anion cases (data not in table). The average distance to the closest water oxygen shows relatively little dependence on ion polarizability, but does show a substantial compression from turning on the ion charge.  The average ion dipole moment magnitude exhibits strong dependence on ion polarizability as expected. As points of reference, the average dipoles for the Cl$^-$ and Br$^-$ ions with their AMOEBA parameters are 1.56 and 2.19 D, respectively, compared with the AIMD values of  0.8 and 1.0 D.\cite{eguar09} It thus appears that the AMOEBA force field is appreciably over-polarizing the ions. The average dipole magnitudes for the waters in the first solvation shell (criterion given above) are slightly suppressed compared to bulk water (2.79 D). Recent AIMD studies have made similar observations for both anions\cite{eguar09} and cations.\cite{whitfield_theoretical_2007}

\section{ Summary}
\label{sec:conc}

This paper has presented an analysis of the thermodynamics and local solvation structure involved in polarizable anion hydration. The theoretical tool employed for the analysis is the quasi-chemical theory of molecular solutions.\cite{ourbook}   The present analysis shows that, in these classical polarizable models, the predominant factors affecting hydration free energies are the ion charge, size, and polarizability, in decreasing order of importance.  Varying the ion size and polarizability independently is physically artificial, but serves to elucidate the relative importance of these factors. The QCT partitions the free energy into three components: inner shell, outer-shell packing, and outer-shell long-ranged contributions.\cite{droge08}   That division is exploited here by noting that, as the conditioning radius is increased, the outer-shell long-ranged contribution assumes a Gaussian character.\cite{shah_balancing_2007,droge08}  The Gaussian nature of the coupling energy distributions in turn implies that a mean-field treatment should be accurate.   This leads to a natural division of the long-ranged component of the hydration free energy into first-order electrostatic, induction, and van der Waals parts.   

We have found that, for the sodium and halide ions examined here, the conditioning radii can be taken as quite small. This is likely due to the spherical nature of these simple ions; future work will focus on application of this approach to molecular ions such as the nitrate ion.  With these small conditioning radii, the $\mu^{\mathrm{ex}}_{\mathrm{IS}} + \mu^{\mathrm{ex}}_{\mathrm{OS,HS}}$ (inner shell plus packing) combination is small and dominated by the packing part. Thus $\mu^{\mathrm{ex}}_{\mathrm{IS}} + \mu^{\mathrm{ex}}_{\mathrm{OS,HS}}$ exhibits ion specificity mainly through the purely size-dependent packing piece.  We also found that the sum of the inner-shell, packing, and long-ranged vdW parts of the free energy is roughly constant, suggesting electrostatic effects (induced by ion size) are the primary determinants of specificity for the chosen hard-sphere conditioning radii.  The inner-shell piece does increase somewhat in magnitude (becomes more negative) with increasing ion polarizability, indicating electrostriction of nearby waters, but this effect is small for the conditioning radii used here.  
The main contributions to the hydration free energy come from the first-order electrostatic and induction terms.  The induction contribution comprises about 25\% of the electrostatic part of the free energy.  The first-order electrostatic piece becomes slightly less favorable with increasing ion polarizability, while the induction piece becomes more favorable.  Due to the induction contribution, the total free energy becomes slightly more favorable with increasing ion polarizability.  

Comparison of the electrostatic potential at the center of nonpolar cavities and ions illustrates substantial differences between the fixed-charge SPC/E and polarizable AMOEBA models.  For the nonpolar cavity case, the electrostatic potential at the cavity center is reduced in the polarizable AMOEBA water model compared to the SPC/E water model.  This occurs even though the average total dipole magnitudes for the nearby waters are larger for the AMOEBA model than for the SPC/E model.  This illustrates the importance of many-body effects in forming the electrostatic environment near the cavity.  Those differences are accentuated when the average potentials are examined for the Na$^+$ and Cl$^-$ ions.  In addition, the SPC/E model exhibits a clear break in the slope of the potential vs.~$\lambda$ plot at radii close to the first maximum of the ion-water(oxygen) rdf, while the polarizable model does not display such a break.  Apparently the many-body polarization smooths out the energy changes as waters are pushed away from the ions.   To examine further the electrostatic contribution to the hydration free energy, we compared the long-ranged electrostatic contribution to the Born model of solvation.  Those results show that, while the Born model fits the simulation data well considering its simplicity, it fails to capture molecular-level details important for accurate computations of free energies.  

Finally, we found that, contrary to the hydration free energies, the local solvation structure is affected more by ion polarizability than by size.  Increasing the ion polarizability leads to more asymmetric solvation environments.  Changes in size for a given polarizability lead to quite small changes in the solvation anisotropy.  The average dipole moment magnitudes for waters in the first solvation shell are slightly suppressed from the bulk value.  This highlights the importance of many-body water-water interactions in responding to the strong field of the nearby ion.  The average ion dipoles, however, displayed values considerably larger (by a factor of 2) than values observed in AIMD quantum simulations.\cite{eguar09}   

\section{ Discussion}
\label{sec:disc}

The results presented here for the hydration free energies show good agreement with experiment\cite{mtiss98,rschm00} and previous simulations employing the AMOEBA model.\cite{agros03}  Based on the free energy calculations alone, the AMOEBA model appears to provide an excellent representation of the hydration of simple alkali halide ion pairs. The quasi-chemical approach allows us to go further, however, in disentangling the various contributions to the free energies.  In particular, the quasi-chemical partitioning has allowed us to examine in detail the relative contributions of ion size and polarizability in hydration.  

Examination of the structural properties of the surrounding solvation shell points out some potential limitations of the AMOEBA model, and likely other models which possess limited or no damping of close-ranged electrostatic interactions.  Namely, as discussed above, the anions are over-polarized compared with AIMD simulations,\cite{dtobi01,sraug02,eguar09} other quantum models,\cite{aohrn04} and recent carefully parametrized classical models.\cite{mmasi08}  In another paper supporting this point, we have examined the local structure in related QM/MM calculations.\cite{zhao09} In that work we generated an ensemble of configurations from an AMOEBA simulation and then performed MP2-level electronic structure calculations on clusters extracted from the ensemble.  The ion plus the nearest 6 waters were treated quantum mechanically, while the more distant waters produced an electrostatic potential due to their AMOEBA-generated multipole moments.  The quantum charge distributions were analyzed with both the ChelpG and QTAIM methods.\cite{szefczyk}  We found that, in agreement with the AMOEBA simulations, the closest waters possess dipoles close to the bulk value due to collective interactions between the water molecules. The anion dipoles are much smaller (roughly 0.6 D for the Cl$^-$ ion) than those predicted by the AMOEBA force field (1.56 D), however.  In addition, we observed substantial charge transfer (of magnitude 0.2 charges) from the anion to the surrounding waters.  This finding is consistent with several previous papers,\cite{thompson_frequency_2000,marenich_polarization_2007,peraro,bucher_polarization_2006}  but differs from the conclusions of Ref.~\cite{rashin_charge_2001} that applied a different charge-accounting scheme. We note that the anion dipole estimate in Ref.~\cite{zhao09} is somewhat smaller than the AIMD prediction; this may be due to the observed charge transfer which is not accounted for in the Wannier analysis typically employed with AIMD simulations.\cite{eguar09} 

The majority of classical simulations that have examined surface affinity of anions have targeted the ion polarizability as the key determinant of the observed segregation.\cite{jungwirth_specific_2006,tchan06}  Other simulations have emphasized the important role of ion size,\cite{beggi08,vaitheeswaran_hydrophobic_2006}  while Hagberg, Brdarski, and Karlstrom\cite{dhagb05} have shown that both factors play important roles. Herce {\it et al.},\cite{sagui05} using polarizable models, suggested that the main causes of surface solvation of a single ion in a water cluster are both
water and ion polarization, coupled to the charge and size of the ion.
On the other hand, Wick and Xantheas\cite{cwick09} have linked anisotropic local solvation structure to surface affinity -- perhaps the ions exhibiting the most asymmetric solvation shells prefer the highly asymmetric liquid-vapor interface region.  
Intuitively, many-body polarization should be a major
contributor,\cite{aohrn07,cwick09} especially when the large electric
field in the region of the water/vapor interface is noted.  
The observed over-polarization in some classical models and charge transfer in quantum studies both suggest that further improvements in the representation of polarization in classical models should be considered to help to quantify the forces driving surface affinity of anions.
A recent study by Wick\cite{wickpol_09} has confirmed that increased damping of the polarization interactions, leading to reduced ion polarization (in agreement with AIMD simulations), in fact results in reduced anion surface affinity. Wick's work is consistent with previous results by Ishiyama and Morita\cite{ishiyama_molecular_2007} that employed reduced polarizability models.  

A first remedy for the anion over-polarization in the classical models would be to reduce the anion polarizability employed until agreement is obtained with {\it ab initio} results for the average dipole magnitudes.  Masia\cite{mmasi08} has argued against this approach, however, since this does not properly represent the physics of the dipoles at close contact; a large magnitude tail can still exist in the distribution of dipole moments  due to over-polarization from closeby waters.  A second remedy would be to re-parametrize the Thole-type damping functions to mimic more closely the longer-ranged damping functions developed by Masia.\cite{mmasi08} A third remedy would be to focus on a more accurate QM/MM approach as outlined above.\cite{zhao09}  We are currently developing such a QM/MM (MP2 level) approach for computing the free energy contributions in the context of quasi-chemical theory.  We have chosen this level of theory so that many-body dispersion interactions as well as accurate representations of local charge rearrangements can be included accurately.  Such studies should help to elucidate further the driving forces for ion surface segregation, and allow for tests of recently developed theories\cite{bostrom_hofmeister_2005,manciu_interactions_2005} of this phenomenon.


\section{Acknowledgements}

  We gratefully acknowledge the support of the NSF (CHE-0709560), the Army MURI program (DAAD19-02-1-0227), and the DOE Computational Science Graduate Fellowship (DE-FG02-97ER25308) for the support of this work. We acknowledge the Ohio Supercomputer Center for a grant of supercomputer time. We thank Zhen Zhao for many helpful discussions concerning her quantum mechanical results.


\begin{thebibliography}{10}

\bibitem{wkunz04}
W.~Kunz, P.~{Lo Nostro}, and B.~W. Ninham,
\newblock Curr. Opin. Colloid Interface Sci. {\bf 9}, 1 (2004).

\bibitem{petrache_salt_2006}
H.~I. Petrache, T.~Zemb, L.~Belloni, and V.~Parsegian,
\newblock Proc. Natl. Acad. Sci. USA {\bf 103}, 7982 (2006).

\bibitem{zhang_interactions_2006}
Y.~J. Zhang and P.~S. Cremer,
\newblock Curr. Op. Chem. Biol. {\bf 10}, 658 (2006).

\bibitem{beggi08}
B.~L. Eggimann and J.~I. Siepmann,
\newblock J. Phys. Chem. C {\bf 112}, 210 (2008).

\bibitem{vaitheeswaran_hydrophobic_2006}
S.~Vaitheeswaran and D.~Thirumalai,
\newblock J. Am. Chem. Soc. {\bf 128}, 13490 (2006).

\bibitem{jungwirth_ions_2002}
P.~Jungwirth and D.~J. Tobias,
\newblock J. Phys. Chem. B {\bf 106}, 6361 (2002).

\bibitem{dhagb05}
D.~Hagberg, S.~Brdarski, and G.~Karlstrom,
\newblock J. Phys. Chem. B {\bf 109}, 4111 (2005).

\bibitem{tchan06}
T.-M. Chang and L.~X. Dang,
\newblock Chem. Rev. {\bf 106}, 1305 (2006).

\bibitem{collins_ions_2007}
K.~D. Collins, G.~W. Neilson, and J.~E. Enderby,
\newblock Biophys. Chem. {\bf 128}, 95 (2007).

\bibitem{pusch_gating_1995}
M.~Pusch, U.~Ludewig, A.~Rehfeldt, and T.~J. Jentsch,
\newblock Nature {\bf 373}, 527 (1995).

\bibitem{zhang_inverse_2009}
Y.~Zhang and P.~Cremer,
\newblock Proc. Natl. Acad. Sci. USA {\bf 106}, 15249 (2009).

\bibitem{vrbka_propensity_2004}
L.~Vrbka et~al.,
\newblock Curr. Op. Coll. \& Int. Sci. {\bf 9}, 67 (2004).

\bibitem{mmuch05}
M.~Mucha et~al.,
\newblock J. Phys. Chem. B {\bf 109}, 7617 (2005).

\bibitem{raymond_probingmolecular_2004}
E.~A. Raymond and G.~L. Richmond,
\newblock J. Phys. Chem. B {\bf 108}, 5051 (2004).

\bibitem{petersen_nature_2006}
P.~B. Petersen and R.~J. Saykally,
\newblock Ann. Rev. Phys. Chem. {\bf 57}, 333 (2006).

\bibitem{ghosal_electron_2005}
S.~Ghosal et~al.,
\newblock Science {\bf 307}, 563 (2005).

\bibitem{jchen06}
J.~Cheng, C.~D. Vecitis, M.~R. Hoffmann, and A.~J. Colussi,
\newblock J. Phys. Chem. B {\bf 110}, 25598 (2006).

\bibitem{jchen08}
J.~Cheng, M.~R. Hoffmann, and A.~J. Colussi,
\newblock J. Phys. Chem. B {\bf 112}, 7157 (2008).

\bibitem{mcari97}
M.~A. Carignano, G.~Karlstrom, and P.~Linse,
\newblock J. Phys. Chem. B {\bf 101}, 1142 (1997).

\bibitem{sagui05}
D.~H. Herce, L.~Perera, T.~A. Darden, and C.~Sagui,
\newblock J. Chem. Phys. {\bf 122}, 024513 (2005).

\bibitem{jungwirth_specific_2006}
P.~Jungwirth and D.~J. Tobias,
\newblock Chem. Rev. {\bf 106}, 1259 (2006).

\bibitem{lpere92}
L.~Perera and M.~L. Berkowitz,
\newblock J. Chem. Phys. {\bf 96}, 8288 (1992).

\bibitem{sstua96}
S.~J. Stuart and B.~J. Berne,
\newblock J. Phys. Chem. {\bf 100}, 11934 (1996).

\bibitem{bostrom_hofmeister_2005}
M.~Bostrom, W.~Kunz, and B.~Ninham,
\newblock Langmuir {\bf 21}, 2619 (2005).

\bibitem{lima_specific_2008}
E.~Lima et~al.,
\newblock J. Phys. Chem. B {\bf 112}, 1580 (2008).

\bibitem{manciu_interactions_2005}
M.~Manciu and E.~Ruckenstein,
\newblock Langmuir {\bf 21}, 11312 (2005).

\bibitem{mcgrath_simulating_2006}
M.~J. {McGrath} et~al.,
\newblock J. Phys. Chem. A {\bf 110}, 640 (2006).

\bibitem{santra_accuracy_2008}
B.~Santra et~al.,
\newblock J. Chem. Phys. {\bf 129}, 194111 (2008).

\bibitem{schmidt_isobaric-isothermal_2009}
J.~Schmidt et~al.,
\newblock J. Phys. Chem. B {\bf 113}, 11959 (2009).

\bibitem{parsons_ab_2009}
D.~F. Parsons and B.~W. Ninham,
\newblock J. Phys. Chem. A {\bf 113}, 1141 (2009).

\bibitem{parsons_nonelectrostatic_2009}
D.~F. Parsons, V.~Deniz, and B.~W. Ninham,
\newblock Colloids Surf., A {\bf 343}, 57 (2009).

\bibitem{barrydrew09}
D.~Parsons and B.~Ninham,
\newblock (2009),
\newblock preprint.

\bibitem{wood_free_1999}
R.~H. Wood, E.~M. Yezdimer, S.~Sakane, J.~A. Barriocanal, and D.~J. Doren,
\newblock J. Chem. Phys. {\bf 110}, 1329 (1999).

\bibitem{sakane_exploringab_2000}
S.~Sakane et~al.,
\newblock J. Chem. Phys. {\bf 113}, 2583 (2000).

\bibitem{liu_hydration_2003}
W.~B. Liu, R.~H. Wood, and D.~J. Doren,
\newblock J. Chem. Phys. {\bf 118}, 2837 (2003).

\bibitem{ituno95}
I.~{Tu{\~n}{\'o}n}, M.~T.~C. Martins-Costa, C.~Millot, and M.~F.
  Ruiz-L{\'o}pez,
\newblock Chem. Phys. Lett. {\bf 241}, 450 (1995).

\bibitem{aohrn04}
A.~{\"O}hrn and G.~Karlstr{\"o}m,
\newblock J. Phys. Chem. B {\bf 108}, 8452 (2004).

\bibitem{leung_ab_2009}
K.~Leung, S.~Rempe, and O.~{von Lilienfeld},
\newblock J. Chem. Phys. {\bf 130}, 204507 (2009).

\bibitem{agros03}
A.~Grossfield, P.~Ren, and J.~W. Ponder,
\newblock J. Am. Chem. Soc. {\bf 125}, 15671 (2003).

\bibitem{pren03}
P.~Ren and J.~W. Ponder,
\newblock J. Phys. Chem. B {\bf 107}, 5933 (2003).

\bibitem{pren04}
P.~Ren and J.~W. Ponder,
\newblock J. Phys. Chem. B {\bf 108}, 13427 (2004).

\bibitem{thalg92}
T.~A. Halgren,
\newblock J. Am. Chem. Soc. {\bf 114}, 7827 (1992).

\bibitem{meng_molecular_1996}
E.~C. Meng and P.~A. Kollman,
\newblock J. Phys. Chem. {\bf 100}, 11460 (1996).

\bibitem{harder_polarizable_2005}
E.~Harder, J.~D. Eaves, A.~Tokmakoff, and B.~J. Berne,
\newblock Proc. Natl. Acad. Sci. USA {\bf 102}, 11611 (2005).

\bibitem{gfano06}
G.~S. Fanourgakis and S.~S. Xantheas,
\newblock J. Phys. Chem. A {\bf 110}, 4100 (2006).

\bibitem{defusco_comparison_2007}
A.~Defusco, D.~P. Schofield, and K.~D. Jordan,
\newblock Mol. Phys. {\bf 105}, 2681 (2007).

\bibitem{glamo06}
G.~Lamoureux and B.~Roux,
\newblock J. Phys. Chem. B {\bf 110}, 3308 (2006).

\bibitem{srick94}
S.~W. Rick, S.~J. Stuart, and B.~J. Berne,
\newblock J. Chem. Phys. {\bf 101}, 6141 (1994).

\bibitem{warren_hydration_2007}
G.~L. Warren and S.~Patel,
\newblock J. Chem. Phys. {\bf 127}, 064509 (2007).

\bibitem{gwarr08}
G.~L. Warren and S.~Patel,
\newblock J. Phys. Chem. C {\bf 112}, 7455 (2008).

\bibitem{mmasi06}
M.~Masia, M.~Probst, and R.~Rey,
\newblock Chem. Phys. Lett. {\bf 420}, 267 (2006).

\bibitem{mmasi08}
M.~Masia,
\newblock J. Chem. Phys. {\bf 128}, 184107 (2008).

\bibitem{rashin_charge_2001}
A.~A. Rashin, I.~A. Topol, G.~J. Tawa, and S.~K. Burt,
\newblock Chem. Phys. Lett. {\bf 335}, 327 (2001).

\bibitem{eguar09}
E.~Gu{\'a}rdia, I.~Skarmoutsos, and M.~Masia,
\newblock J. Chem. Theor. Comput. {\bf 5}, 1449 (2009).

\bibitem{cwick09}
C.~D. Wick and S.~S. Xantheas,
\newblock J. Phys. Chem. B {\bf 113}, 4141 (2009).

\bibitem{ishiyama_molecular_2007}
T.~Ishiyama and A.~Morita,
\newblock J. Phys. Chem. C {\bf 111}, 721 (2007).

\bibitem{ourbook}
T.~L. Beck, M.~E. Paulaitis, and L.~R. Pratt,
\newblock {\em The Potential Distribution Theorem and Models of Molecular
  Solutions},
\newblock Cambridge, New York, 2006.

\bibitem{asthagiri_absolute_2003}
D.~Asthagiri, L.~R. Pratt, and H.~S. Ashbaugh,
\newblock J. Chem. Phys. {\bf 119}, 2702 (2003).

\bibitem{asthagiri_quasi-chemical_2003}
D.~Asthagiri and L.~R. Pratt,
\newblock Chem. Phys. Lett. {\bf 371}, 613 (2003).

\bibitem{shah_balancing_2007}
J.~K. Shah, D.~Asthagiri, L.~R. Pratt, and M.~E. Paulaitis,
\newblock J. Chem. Phys. {\bf 127}, 144508 (2007).

\bibitem{merchant_thermodynamically_2009}
S.~Merchant and D.~Asthagiri,
\newblock J. Chem. Phys. {\bf 130}, 195102 (2009).

\bibitem{varma_structural_2008}
S.~Varma and S.~B. Rempe,
\newblock J. Am. Chem. Soc. {\bf 130}, 15405 (2008).

\bibitem{droge08}
D.~M. Rogers and T.~L. Beck,
\newblock J. Chem. Phys. {\bf 129}, 134505 (2008).

\bibitem{sabo_studies_2008}
D.~Sabo, S.~Varma, M.~Martin, and S.~Rempe,
\newblock J. Phys. Chem. B {\bf 112}, 867 (2008).

\bibitem{amber10}
D.~Case et~al.,
\newblock {\em {AMBER 10}},
\newblock University of California, San Francisco, 2008.

\bibitem{laage_reorientional_2007}
D.~Laage and J.~Hynes,
\newblock Proc. Natl. Acad. Sci USA {\bf 104}, 11167 (2007).

\bibitem{ghumm96}
G.~Hummer, L.~R. Pratt, and A.~E. Garcia,
\newblock J. Phys. Chem. {\bf 100}, 1206 (1996).

\bibitem{sraug02}
S.~Raugei and M.~L. Klein,
\newblock J. Chem. Phys. {\bf 116}, 196 (2002).

\bibitem{mtiss98}
M.~D. Tissandier et~al.,
\newblock J. Phys. Chem. A {\bf 102}, 7787 (1998).

\bibitem{rschm00}
R.~Schmid, A.~M. Miah, and V.~N. Sapunov,
\newblock Phys. Chem. Chem. Phys. {\bf 2}, 97 (2000).

\bibitem{hfried73}
H.~L. Friedman and C.~V. Krishnan,
\newblock Thermodynamics of ion hydration,
\newblock in {\em Water: A Comprehensive Treatise}, edited by F.~Franks, Plenum
  Press, New York, 1973.

\bibitem{bkrum96}
B.~S. Krumgalz, R.~Pogorelsky, and K.~S. Pitzer,
\newblock J. Phys. Chem. Ref. Data {\bf 25}, 663 (1996).

\bibitem{topol_structure_1999}
I.~A. Topol, G.~J. Tawa, S.~K. Burt, and A.~A. Rashin,
\newblock J. Chem. Phys. {\bf 111}, 10998 (1999).

\bibitem{srame08}
S.~G. Ramesh, S.~Re, and J.~T. Hynes,
\newblock J. Phys. Chem. A {\bf 112}, 3391 (2008).

\bibitem{stone}
A.~J. Stone,
\newblock {\em The Theory of Intermolecular Forces},
\newblock Oxford University Press, Oxford, 1997.

\bibitem{amisq05}
A.~J. Misquitta, R.~Podeszwa, B.~Jeziorski, and K.~Szalewicz,
\newblock J. Chem. Phys. {\bf 123}, 214103 (2005).

\bibitem{bukowski_predictions_2007}
R.~Bukowski, K.~Szalewicz, G.~C. Groenenboom, and A.~{van der Avoird},
\newblock Science {\bf 315}, 1249 (2007).

\bibitem{agros05}
A.~Grossfield,
\newblock J. Chem. Phys. {\bf 122}, 024506 (2005).

\bibitem{hashb00}
H.~S. Ashbaugh,
\newblock J. Phys. Chem. B {\bf 104}, 7235 (2000).

\bibitem{pyang08}
P.-K. Yang and C.~Lim,
\newblock J. Phys. Chem. B {\bf 112}, 14863 (2008).

\bibitem{skalko96}
S.~G. Kaldo, G.~Sese, and J.~A. Padro,
\newblock J. Chem. Phys. {\bf 104}, 9578 (1996).

\bibitem{whitfield_theoretical_2007}
T.~W. Whitfield et~al.,
\newblock J. Chem. Theor. Comput. {\bf 3}, 2068 (2007).

\bibitem{dtobi01}
D.~J. Tobias, P.~Jungwirth, and M.~Parrinello,
\newblock J. Chem. Phys. {\bf 114}, 7036 (2001).

\bibitem{zhao09}
Z.~Zhao, D.~M. Rogers, and T.~L. Beck,
\newblock J. Chem. Phys.  (2009),
\newblock Submitted. Available at http://arxiv.org/abs/0910.1791.

\bibitem{szefczyk}
B.~Szefczyk, W.~A. Sokalski, and J.~Leszczynski,
\newblock J. Chem. Phys. {\bf 117}, 6952 (2002).

\bibitem{thompson_frequency_2000}
W.~H. Thompson and J.~T. Hynes,
\newblock J. Am. Chem. Soc. {\bf 122}, 6278 (2000).

\bibitem{marenich_polarization_2007}
A.~V. Marenich, R.~M. Olson, A.~C. Chamberlin, C.~J. Cramer, and D.~G. Truhlar,
\newblock J. Chem. Theor. Comput. {\bf 3}, 2055 (2007).

\bibitem{peraro}
M.~{Dal Peraro}, S.~Raugei, P.~Carloni, and M.~L. Klein,
\newblock ChemPhysChem {\bf 6}, 1715 (2005).

\bibitem{bucher_polarization_2006}
D.~Bucher et~al.,
\newblock Biophys. Chem. {\bf 124}, 292 (2006).

\bibitem{aohrn07}
A.~{\"O}hrn and G.~Karlstr{\"o}m,
\newblock J. Chem. Theor. Comput. {\bf 3}, 1993 (2007).

\bibitem{wickpol_09}
C.~D. Wick,
\newblock J. Chem. Phys. {\bf 131}, 084715 (2009).

\end{thebibliography}

\end{document}